\documentclass[11pt,draftclsnofoot,journal,letterpaper,onecolumn]{IEEEtran}
\IEEEoverridecommandlockouts
\usepackage[noadjust]{cite}

\ifCLASSINFOpdf
  \usepackage[pdftex]{graphicx}
\else
\fi
\usepackage[cmex10]{amsmath}
\usepackage{amssymb}   
\usepackage{url}

\usepackage[utf8]{inputenc} 
\usepackage[T1]{fontenc}

\usepackage{mleftright}       
\mleftright                   

\usepackage{booktabs}         

\newtheorem{theorem}{Theorem}
\newtheorem{proposition}[theorem]{Proposition}

\newtheorem{definition}[theorem]{Definition}

\newtheorem{lemma}[theorem]{Lemma}
\newtheorem{claim}[theorem]{Claim}

\newtheorem{remark}[theorem]{Remark}

\newcommand{\iid}{i.i.d.\hspace{2pt}}

\begin{document}
%

\title{The Rate-Distortion-Perception Trade-Off with Algorithmic Realism\\
\thanks{Y. Hamdi and D. G\"{u}nd\"{u}z are with the Department of Electrical and Electronic Engineering, Imperial College London, UK. Emails: \{y.hamdi, d.gunduz\}@imperial.ac.uk. A. B. Wagner is with the School of Electrical and Computer Engineering, Cornell University, USA. Email: wagner@cornell.edu}
\thanks{
We thank Lucas Theis for introducing us to universal critics and for useful discussions and suggestions on how to improve our manuscript.
Part of this paper was presented at the 2025 IEEE International Symposium on Information Theory.
This research was supported by the European Union’s Horizon 2020 program through Marie Skłodowska Curie Innovative Training Network Greenedge (No. 953775), by the United Kingdom Engineering and Physical Sciences Research Council (EPSRC) for the projects AIR (ERC Consolidator Grant, EP/X030806/1) and INFORMED-AI (EP/Y028732/1), and by the US National Science Foundation under grant CCF-2306278. For the purpose of open access, the authors have applied a Creative Commons Attribution (CCBY) license to any Author Accepted Manuscript version arising from this submission.
}
} 
%
%
%


\author{%
  \IEEEauthorblockN{Yassine Hamdi,~\IEEEmembership{Graduate Student Member,~IEEE},
  Aaron B. Wagner,~\IEEEmembership{Fellow,~IEEE},\\
  Deniz G\"{u}nd\"{u}z,~\IEEEmembership{Fellow,~IEEE}
  }
}

\maketitle

\begin{abstract}

Realism constraints (or constraints on perceptual quality) have received considerable recent attention within the context of lossy compression, particularly
of images. Theoretical studies of lossy compression indicate that high-rate common randomness between the compressor and the decompressor is a valuable resource for achieving realism.
On the other hand,
the utility of significant amounts of common randomness
has not been noted in practice.
We offer
an
explanation for this discrepancy
by considering a
realism constraint
that
requires satisfying a universal critic that inspects realizations of individual
compressed
reconstructions,
or batches thereof.
We characterize the optimal
rate-distortion
trade-off under such a realism constraint, and show that it
is asymptotically achievable without any common randomness, unless the batch size is impractically large.
\end{abstract}



%

\section{Introduction}\label{sec:intro}
%
%
%
%
\IEEEPARstart{A}{persistent} open challenge in lossy compression, especially for image/video compression~\cite{Eckert:SP:98, Wu:ICMEW:12}, is the realism, or perceived quality, of reconstructed signals. The tremendous advancements in neural compression methods and image generation models in recent years have sparked a renewed interest in it. Reconstructed images are expected to have a high pixel-level fidelity to the original source and be indistinguishable to humans from naturally occurring ones. This guarantees that there are no visible artifacts, like blocking or blurriness, in the reconstructed images.

The desirability of statistical similarity between the source and reconstruction has long been noted. For example,
MPEG
Advanced Audio Coding (AAC) allows the output to have high-frequency noise added to it~\cite[Sec.~17.4.2]{Sayood:Compression}, in order for its power spectrum to be close to that of the source.
The advent of adversarial loss functions in learnt compression has brought the concept back into the spotlight~\cite{2018FirstGenerativeCompression,2018NeuripsTschannenAgustssonNeedForStchasticInputToGeneratorInCompression,2019AgustssonMentzerGANforExtremeCompression,2019BlauMichaeliRethinkingLossyCompressionTheRDPTradeoff}.
This has been shown to be an effective technique in practice for guaranteeing that reconstructed images have excellent perceptual quality~\cite{2019AgustssonMentzerGANforExtremeCompression,2020MentzerHIFICGenerativeCompression,2022PoELIC,2024CVFWinterbyIwaiEtAlGANbasedRDPwithAdaptiveRate}.
A connection has been established between the minimization of adversarial loss functions and known variational expressions of statistical divergences~\cite{2014GoodfellowGANs,2017WassersteinGAN,2016GANwithAnyDivergence}.
Thus, in addition to demanding that each reconstructed image be close to its individual source according to standard ideas of distortion, one considers
restricting the reconstruction distribution to be close to that of the source according to some divergence.

Several information-theoretic formulations along these lines have been proposed~\cite{1991MomentPreservingQuantization,2010LiEtAlTheirFirstPaperOnDistributionPreservingQuantization,2011LiEtAlMainPaperOnDistributionPreservingQuantization,2012LiEtAlSpectralDensityPreservingQuantizationForAudio,2013LiEtAlMultipleDescriptionDistributionPreservingQuantization,Jan2015SaldiEtAlDistributionPreservationMeasureTheoreticConsiderationsForContinuousAndDiscreteCommonRandomness,Sep2015RDPLimitedCommonRandomnessSaldi,2019BlauMichaeliRethinkingLossyCompressionTheRDPTradeoff,TheisWagner2021VariableRateRDP, Dec2022WeakAndStrongPerceptionConstraintsAndRandomness,SalehkalaibarISIT2024RDPConditioningOnTheMessage}.
Let $X_{1:n} := (X_1, \cdots, X_n)$ denote a source sequence of independent and identically distributed
(i.i.d.) symbols, each distributed according to
$p_X$ on some alphabet $\mathcal{X}.$ The distribution of $X_{1:n}$ is denoted $p_X^{\otimes n}.$ 
Consider an encoder having access to a source sequence $X_{1:n},$ which can communicate with a decoder through a noiseless rate-limited channel.
Let $Y_{1:n}$ denote the reconstruction sequence produced by the decoder on alphabet $\mathcal{X}.$
Two main categories of asymptotic realism constraints have been considered
in the literature,
in conjunction with a traditional additive distortion constraint:
the main constraints in each category
are
the \textit{near-perfect strong realism} constraint
which
imposes
\begin{IEEEeqnarray}{c}
\mathcal{D}(P^{(n)}_{Y_{1:n}}, p_{X}^{\otimes n}) \underset{n\to\infty}{\longrightarrow} 0,
\label{eq:def_strong_realism_constraint}
\IEEEeqnarraynumspace
\end{IEEEeqnarray}
and
the \textit{near-perfect per-symbol realism} constraint
which
requires
\begin{IEEEeqnarray}{c}
\sup_{1\leq t \leq n} \mathcal{D}(P^{(n)}_{Y_t}, p_{X}) \underset{n\to\infty}{\longrightarrow} 0,
\label{eq:def_per_symbol_realism_constraint}
\IEEEeqnarraynumspace
\end{IEEEeqnarray}
where $\mathcal{D}$ is a divergence between distributions.
Imposing \eqref{eq:def_strong_realism_constraint} or \eqref{eq:def_per_symbol_realism_constraint} in addition
to the usual distortion constraint naturally gives rise to the rate-distortion-perception
(RDP)
function~\cite{2011LiEtAlMainPaperOnDistributionPreservingQuantization,Jan2015SaldiEtAlDistributionPreservationMeasureTheoreticConsiderationsForContinuousAndDiscreteCommonRandomness,Sep2015RDPLimitedCommonRandomnessSaldi,TheisWagner2021VariableRateRDP}:
\begin{IEEEeqnarray}{c}
R^{(1)}(\Delta) := \min_{\substack{p_{Y|X} \text{ s.t. } \\
p_X = p_Y
,
\\ \mathbb{E}_
p
[d(X,Y)] \leq \Delta}} I_
p
(X;Y),
\label{eq:def_RDP_function}
\end{IEEEeqnarray}
where $d:\mathcal{X}\times\mathcal{X} \to [0,\infty)$ is the single-letter distortion function.
The emerging consensus~\cite{TheisWagner2021VariableRateRDP,UniversalRDPNeurips2021,Dec2022WeakAndStrongPerceptionConstraintsAndRandomness,SalehkalaibarKhistiNeurips2023RDPVideoCompressionChoiceOfPerceptionLoss,2025JunChenRDPWithDifferentTargetDistribAndIidAssumption,2025TITversionOfUniversalRDPwithPerSymbolRealismAndSuccessiveRefinement} that realism constraints are captured by \eqref{eq:def_strong_realism_constraint} or \eqref{eq:def_per_symbol_realism_constraint}, with the resulting fundamental limit being the RDP function --- or slight variations of \eqref{eq:def_strong_realism_constraint}, \eqref{eq:def_per_symbol_realism_constraint}, and $R^{(1)}$ ---, has two significant limitations, which are the focus of this paper.

First, the goal of a realism constraint is to ensure that the reconstructed image appears realistic to a human observer. Yet  \eqref{eq:def_strong_realism_constraint} and \eqref{eq:def_per_symbol_realism_constraint} are arguably mismatched to human perception,
as already noted after Remark 11 in \cite{Dec2022WeakAndStrongPerceptionConstraintsAndRandomness}.
Under \eqref{eq:def_strong_realism_constraint} and \eqref{eq:def_per_symbol_realism_constraint}, it is not meaningful to speak of
whether an individual realization is realistic; realism is a property of the ensemble. While human vision can quickly identify unrealistic 
aspects of individual images, it is blind to many of the ensemble-level deviations that are central to \eqref{eq:def_strong_realism_constraint} and \eqref{eq:def_per_symbol_realism_constraint}. Humans would not be able to distinguish the ensemble of natural images from those
selected at random from a very large but fixed set, for instance, nor would they be able to tell, by sight alone, if an ensemble of images is 
slightly more, or slightly less, compressible under a given algorithm such as JPEG than the ensemble of natural images.

Second, theoretical studies that adopt \eqref{eq:def_strong_realism_constraint}
as realism
constraint
indicate that large amounts of high-quality common randomness between the encoder and the decoder are required in order
to achieve the RDP
function~\cite{Sep2015RDPLimitedCommonRandomnessSaldi}.\footnote{It is conjectured that the same can be said regarding
variants of \eqref{eq:def_strong_realism_constraint}
of the form
$
\limsup_{n\to\infty}
\mathcal{D}(P^{(n)}_{Y_{1:n}}, p_{X}^{\otimes n})
\leq \lambda
,$
where $\lambda$ is a real number
--- although this was only proved \cite[Remark~11]{Dec2022WeakAndStrongPerceptionConstraintsAndRandomness} under the assumption that the reconstruction $Y_{1:n}$ is i.i.d..}\textsuperscript{,}\footnote{Note that these results indicate that pseudorandomness is not sufficient;
thus the encoder cannot simply communicate a low-rate seed to the decoder.} Yet no work has experimentally observed the need for a large
amount of common randomness. Indeed, it would be surprising if a practical scheme ceased producing realistic images when it was derandomized by
replacing the common randomness with a fixed realization.

We argue in favor of replacing \eqref{eq:def_strong_realism_constraint} and \eqref{eq:def_per_symbol_realism_constraint} with a realization-based
realism constraint. 
Specifically, we consider a \emph{critic} that assigns a realism deficiency (or atypicality) score to
each reconstruction. We then require that the expected deficiency score is asymptotically bounded.
This is more aligned with human perception and, as we shall see, it eliminates the need for common randomness for
all practical purposes.
Theoretical frameworks involving
realization-based measures of realism
have been considered previously~\cite{Sep2015RDPLimitedCommonRandomnessSaldi,Dec2022WeakAndStrongPerceptionConstraintsAndRandomness}, although they were restricted to comparing the empirical
distribution of the source and the reconstruction, or merely requiring that the reconstruction lie in the support of the source distribution.
Specifically, if 
$\hat{P}_{y_{1:n}}$ is the empirical distribution of a realization $y_{1:n}$ of the reconstruction, then we could use
$\mathcal{D}(\hat{P}_{y_{1:n}}, p_X)$ as
the critic.
This approach, as well as adopting \eqref{eq:def_per_symbol_realism_constraint}, would fail to impose any constraint on the order of the symbols in the reconstruction sequence
$y_{1:n}.$
For instance,
the reconstruction $101010 \cdots$ would be deemed
a realistic reconstruction if the source is Bernoulli($1/2$), when in fact this realization is atypical in all but
the weakest sense.

Inspired by Theis~\cite{2024TheisUniversalCriticsPositionPaper},
we impose a stronger constraint in which we require that every computable critic
is bounded. Any reconstruction that meets such a constraint must be typical in a strong sense.
It must asymptotically have the correct $k$th-order marginals for any fixed $k$ (see Proposition~\ref{prop:frequency_critic_any_alphabet} to follow). 
But it also must be consistent with other, more subtle limit theorems, such as the Erd\H{o}s-R\'{e}nyi law of runs and
its generalizations: for an $n$-length i.i.d.\ sequence of Bernoulli($p$) random variables, with $n$ large,
the length of the longest run of ones is within a small additive gap of $\log_{1/p} n$ with high 
probability~\cite{GordonRunsPTRF,GuibasLongPatternsZWVG,BoydLosingRunsUnpublished}.
The concentration around $\log_{1/p} n$ is sufficiently tight that it can be used to distinguish truly
random bits from a human-generated sequence; the runs in the latter are inevitably too short~\cite{SchillingRunCollegeMath,Revesz1978StrongTheoremsICM}. 
A reconstruction that has small realism deficiency 
under every computable critic must have a longest run that is close to
$\log_{1/p} n$. It must likewise satisfy every other limit theorem satisfied by the source 
distribution, including those not yet discovered. 

Although such a realism constraint is very strong, we show that it is essentially free: it is satisfied
with high probability by a compression scheme constructed through random selection in the usual fashion.
In particular, under this realism constraint, the RDP function in \eqref{eq:def_RDP_function}
can be achieved with deterministic codes. The idea is that each codeword is realistic according
to a given critic because it is generated according to the source distribution. To argue that it is realistic
according to all computable critics, we use the fact that the number of such critics is countable,
and thus, we can create a ``universal'' critic whose realism deficiency is
approximately
a convex combination of all
computable deficiencies.
If the expected deficiency of this universal critic is asymptotically bounded then
it must be bounded for each of the constituent critics.

The universal critic
$\delta: \cup_{n\in\mathbb{N}} \mathcal{X}^n \to \mathbb{R}$
is itself incomputable
and can be shown to
be related to
the Kolmogorov complexity
$x_{1:n} \mapsto K(x_{1:n})$
(see Remark \ref{remark:ties_to_Kolmogorov_complexity}
to follow and Appendix \ref{app:proof_of_existence_universal_critic}).
Using the notion of Kolmogorov complexity in the formulation of an engineering problem inevitably raises the objection
that $K(x_{1:n})$ is not computable~\cite[Sec.~14.7]{Cover&Thomas2006}. This is not a concern in the present application because it is the critic, not
the compression scheme, that is called upon to compute $K(x_{1:n})$. By showing that a code can achieve the
RDP
function without common randomness while satisfying this most powerful of
critics, we show that it can also satisfy any weaker critic, such as a human observer. This is akin to
how proofs of information-theoretic security assume adversaries with unlimited computational power.

The realization-based approach to realism espoused in this paper might appear to be of a fundamentally
different nature from the divergence-based approach in \eqref{eq:def_strong_realism_constraint} and \eqref{eq:def_per_symbol_realism_constraint}.
In fact, it is possible to simultaneously generalize the two by extending the framework to
allow the critic to examine batches of reconstructions at a time~\cite{2024TheisUniversalCriticsPositionPaper}. Specifically, when the blocklength is
$n$, we assume that $B_n$ source realizations are generated and passed through the encoder and
decoder. The critic then examines the resulting $B_n$ reconstructions. For $B_n = 1$, this reduces to
our original formulation, and no common randomness is needed to achieve the RDP
function. On the other hand, if $B_n$ grows very quickly with $n$, then the batch essentially reveals
the $n$-length distribution of the reconstructions, and the realism constraint is equivalent to the
divergence constraint in \eqref{eq:def_strong_realism_constraint}. In this case, common randomness
is needed to achieve the RDP
function.
More generally, we show that if $B_n$ grows sub-exponentially with $n$, then common randomness is not
needed to achieve the RDP
function while it is necessary if $B_n$ grows
exponentially with $n$ at a sufficiently large rate. Characterizing the extent to which 
common randomness is needed in the intermediate regime between these two extremes is
an interesting theoretical problem raised by this work.
In the case of sub-exponential $B_n$,
these findings shed light on why the need for common randomness has not been observed in practice:
to satisfy a realization-based critic such as a human-observer, common randomness is only needed
if the critic considers impractically large batches at a time.

In Section \ref{sec:background},
we provide
some background on the
formalism for critics in algorithmic information theory.
In Section \ref{sec:new_RDP_formalism_and_all_results},
we introduce our new formalism for the RDP trade-off.
In Section \ref{sec:statements_of_3_main_theorems}, we state our main results, namely Theorems \ref{thm:small_batch_size_asymptotics}, \ref{thm:one_shot_achievable_points}, and \ref{thm:very_large_batch_size}.
All
proofs are deferred to the appendices.

\section{Background}\label{sec:background}

\subsection{Notation}
Calligraphic letters such as $\mathcal{X}$ denote sets, except in $p^{\mathcal{U}}_{\mathcal{J}},$ which denotes the uniform distribution over the set $\mathcal{J}.$
The cardinality of a finite set $\mathcal{X}$ is denoted by $|\mathcal{X}|.$
Random variables are denoted by upper case letters such as $X$ and their realizations by lower case letters such as $x.$
For a distribution $P,$ the expression $P_X$ denotes the marginal of variable $X$ while $P(x)$ denotes a real number, probability of the event $X=x.$ Similarly $P_{X|Y=y}$ denotes a distribution over $\mathcal{X}$ and $P_{X|Y=y}(x)$ the corresponding real number.
Given a real number $\tau,$ we denote by $\lfloor \tau \rfloor$ (resp. $\lceil \tau \rceil$) the largest (resp. smallest) integer less (resp. greater) than or equal to $\tau.$
We denote by $[\tau]$ the set $\{1, ..., \lfloor \tau \rfloor\}$ and by $\{0,1\}^*$ the set of non-empty finite strings of $0$'s and $1$'s.
We use $x_{1:n}$ to denote a finite sequence $(x_1, ..., x_n),$ and $\mathbf{x}^{(n,b)}$ to denote a batch $\{x^{(k)}_{1{:}n}\}_{k {\in} [b]}$ of $b$ strings, each
of length $n.$
We abbreviate $\mathbf{x}^{(1,b)}$ with $\mathbf{x}^{(b)}.$
When the alphabet of a sequence $x$ is clear from the context, we denote the length
of the latter
by $l(x),$ e.g., $l(x_{1:n})=n.$
We denote the set of (strictly) positive reals by $\mathbb{R}_+,$ the set of (strictly) positive integers by $\mathbb{N},$ and the set of rational numbers by $\mathbb{Q}.$
The closure of a set
$\mathcal{A}$
is denoted by
$cl(\mathcal{A}).$
The notation $:=$ stands for ``is defined by", and $\equiv$ denotes equality of distribution.
We use
$I_p(X;Y)$ to denote the mutual information between random variables $X$ and $Y$ with respect to joint distribution $p_{X,Y}.$
Logarithms are in base 2.
The
total variation distance (TVD)
between distributions $p$ and $q$ on a finite set $\mathcal{X}$ is defined by
$
\|p-q\|_{TV} := \tfrac{1}{2} \sum_{x\in\mathcal{X}} |p(x)-q(x)|.
$
For any nonempty finite set $\mathcal{X},$
and any distribution $p$ on $\mathcal{X},$ we denote by $p^{\otimes *}$ the function defined on
$\cup_{n\in \mathbb{N}} \mathcal{X}^n,$ and such that for every $n \in \mathbb{N},$ the restriction of $p^{\otimes *}$ on $\mathcal{X}^n$ is $p^{\otimes n}.$
For a finite set 
$\mathcal{X},$
the empirical distribution of
a sequence
$x_{1:n}{\in}\mathcal{X}^n$ is denoted $\mathbb{P}^{\text{emp}}_{\mathcal{X}}(x_{1:n}).$
Given a distribution $P_{X_{1:n}}$ on
$\mathcal{X}^n,$ we denote by $\hat{P}_{\mathcal{X}}[X_{1:n}]$
the \textit{average marginal distribution} of random string $X_{1:n},$ i.e.,
$
\hat{P}_{\mathcal{X}}[X_{1:n}]
:= \tfrac{1}{n} \sum_{t=1}^n P_{X_t}
.
$

\subsection{Background on algorithmic information theory}
The theory of $p$-critics and universal 
critics
has recently been brought to the attention of
the machine
vision
community via \cite{2024TheisUniversalCriticsPositionPaper}. We refer to it for
a high-level and insightful presentation of the topic and its usefulness in diverse machine learning tasks such as generative modeling and outlier detection.
Relevant
background
on computability theory is
provided
in Appendix \ref{app:algorithmic_information_theory}.
Throughout the paper, we assume that the source
distribution $p_X$
is
a computable function
from
a finite set $\mathcal{X}$
to $(0,1).$
Algorithmic information theoretic notions are commonly defined for binary strings.
These can be extended to strings of elements of $\mathcal{X}$ by fixing
an injection from $\mathcal{X}$ to $\{0,1\}^s,$ for some $s\in\mathbb{N}.$
\begin{definition}\label{def:sum_critic_at_beginning_of_paper}
Consider
a finite set $\mathcal{X}.$
Let
$p$ be a distribution on $\mathcal{X}$ such that $\forall x \in \mathcal{X}, p(x)>0.$ 
A \textit{$p$-critic} is a
function $\delta: 
\mathcal{X}
\to \mathbb{R},$ such that
\begin{IEEEeqnarray}{c}
\sum_{x\in
\mathcal{X}
} p(x
)2^{\delta(x)} \leq 1.
\label{eq:bound_in_def_sum_critic_if_supported_on_fixed_length_set}
\end{IEEEeqnarray}
A
$p^{\otimes *}$
-critic is a
function $\delta: \cup_{n\in\mathbb{N}} \mathcal{X}^n \to \mathbb{R},$ such that
for every input dimension $n\in\mathbb{N},$
we have
\begin{IEEEeqnarray}{c}
\sum_{x\in
\mathcal{X}
^n
}
p^{\otimes n}
(x
)2^{\delta(x)} \leq 1.
\label{eq:bound_in_def_sum_critic_n_letter}
\end{IEEEeqnarray}
\end{definition}

The
notion of $
p^{\otimes *}
$-critic in Definition \ref{def:sum_critic_at_beginning_of_paper} will be used
to study
asymptotic regimes
in Section \ref{sec:formulation_asymptotic_achievability}.
For any $n\in\mathbb{N},$ let $\mathbf{1}_{\mathcal{X}^n}:\cup_{n\in\mathbb{N}} \mathcal{X}^n \to \{0,1\}$ be such that $\mathbf{1}_{\mathcal{X}^n}(x)=1$ iff $x\in\mathcal{X}^n.$
Note that for any probability distribution $\pi$ on $\mathbb{N},$ the mixture $\Tilde{p}:=\sum_{n\in\mathbb{N}}\pi(n)\mathbf{1}_{\mathcal{X}^n}p^{\otimes n}$ is a probability measure. By multiplying \eqref{eq:bound_in_def_sum_critic_n_letter} by
$
\pi(n)
,$ and summing over $n,$ we obtain
\begin{IEEEeqnarray}{c}
\sum_{x\in
\cup_{n\in\mathbb{N}} \mathcal{X}^n
} \Tilde{p}(x)2^{\delta(x)} \leq 1.
\label{eq:conventional_sum_test_requirement}
\end{IEEEeqnarray}
Thus a $p^{\otimes *}$-critic can be viewed as a critic over $\mathcal{X}^*$.

A critic is essentially
a log-likelihood ratio: given a $p$-critic
$\delta,$ setting $q: x \mapsto p(x)2^{\delta(x)}$
gives
\begin{IEEEeqnarray}{c}
\delta(x) = \log\Big(\dfrac{q(x)}{p(x)}\Big)
\text{ and }
\
\sum_{x\in\mathcal{X}} q(x) \leq 1.
\label{eq:sum_critic_is_a_likelihood_ratio}
\end{IEEEeqnarray}
Conversely, for any sequence of sub-probability distributions\footnote{A real-valued function $q$ defined on a set $\mathcal{E}$ is a sub-probability distribution if $\forall x \in \mathcal{E}, q(x)\geq 0,$ and $\sum_{x\in\mathcal{E}} q(x) \leq 1.$} $\{q_n\}_{n\in\mathbb{N}}$, with the $n$-th
defined on $\mathcal{X}^n,$
\begin{IEEEeqnarray}{c}
x_{1:n} \mapsto \log \frac{q_n(x_{1:n})}{p_X^{\otimes n}(x_{1:n})}
\label{eq:any_LLR_is_a_critic}
\end{IEEEeqnarray}
is a $p^{\otimes *}$-critic.
Setting $q_n = q^{\otimes n}$ in \eqref{eq:any_LLR_is_a_critic} for some distribution $q$
yields the $p^{\otimes *}$-critic
\begin{IEEEeqnarray}{c}
   \delta(x_{1:n}) = \log \frac{q^{\otimes n}(x_{1:n})}{p_X^{\otimes n}(x_{1:n})} = \sum_{i = 1}^n \log \frac{q(x_i)}{p_X(x_i)}.
   \label{eq:critic_lln}
\end{IEEEeqnarray}
If $X_{1:n}$ is indeed i.i.d.\ with $p_X$, then $\delta(X_{1:n})$ tends linearly to negative 
infinity, while if it is actually
i.i.d.\ with $q$, then it tends linearly to infinity. Such a test detects 
``large'' deviations in the empirical distribution of the source. With additional effort, we can
construct critics that test for smaller-scale deviations.

\begin{proposition}\label{prop:frequency_critic_any_alphabet}
Consider
a finite set $\mathcal{X}.$
Let
$q$ be a distribution on $\mathcal{X}$ such that $\forall x \in \mathcal{X}, q(x)>0.$ 
Let $e_0$ be any symbol in $
\mathcal{X}
.$
For any $n\in\mathbb{N}$ and any $x_{1:n
}\in
\mathcal{X}
^n
,$ let $S(x_{1:n
})$ denote the number of occurrences of $e_0$ in
$x_{1:n
}.$
Define map $\delta: \cup_{n\in\mathbb{N}} \mathcal{X}^n \to \mathbb{N}_{\geq 0}$ by: $\forall n \in\mathbb{N}, \forall x_{1:n} \in \mathcal{X}^n,$
\begin{IEEEeqnarray}{c}
x
_{1:n}
\mapsto
\Big
\lceil
\log\Big\lceil
|S(x
_{1:n}
)-q(e_0)
n
| \ \big/ \ \sqrt{
n
}
\Big\rceil \Big
\rceil
\nonumber
\end{IEEEeqnarray}
if $S(x_{1:n})\neq q(e_0)n,$
and $x_{1:n} \mapsto 0$ otherwise.
Then, $\delta-2\log(\delta+
3
)
$ is a computable $q^{\otimes *}$-critic.
\end{proposition}

The proof is provided in Appendix~\ref{app:frequency_critic}. If $X_{1:n}$ is indeed i.i.d.\ with $p_X$, then
this critic's output will remain bounded in expectation as $n \rightarrow \infty$. But if its distribution is such that
the empirical fraction of $e_0$ symbols departs from $q(e_0)$ by, say, $1/n^{1/2-\epsilon}$, then the
output will grow without bound. 

The critic in \eqref{eq:critic_lln} is in some sense testing whether
a realization fails to satisfy the prediction of the law of large numbers. The critic
in Proposition~\ref{prop:frequency_critic_any_alphabet}
is testing a form of the central limit theorem. More generally, any
theorem showing that source realizations must satisfy a certain property with high probability 
can potentially be embodied in a critic. We illustrate this with the Erd\H{o}s-R\'{e}nyi 
theory of runs.

\begin{proposition}
\label{prop:critic_longest_run}
Suppose $\mathcal{X} = \{0,1\}$ and $p$ is Bernoulli($q$), where $q\in (0,1/2]$
is a computable real number.
For any $n\in\mathbb{N}$ and $x_{1:n}\in\mathcal{X}^n,$
let $R(x_{1:n})$ denote the length of the longest run of ones in $x_{1:n}.$ Then there exists
a bounded, computable sequence $\{a_n\}_{n\in\mathbb{N}}$ of real numbers, such that
the function
\begin{IEEEeqnarray}{c}
\delta_{\textrm{run}}
: \cup_{n\in\mathbb{N}}\mathcal{X}^n \to \mathbb{R}, \ \
x_{1:n}
\mapsto
\log\left(1+|R(x_{1:n}) - \log_{1/q} n|\right) - a_n
\nonumber
\end{IEEEeqnarray}
is a computable $p^{\otimes *}$-critic.
\end{proposition}

The proof is provided in Appendix \ref{app:subsec:proof_Erdos_Renyi}.
If $X_{1:n}$ is i.i.d.\ Bernoulli($r$), where $r \ne q$, then $R(X_{1:n})$ will grow as
$\log_{1/r} n$, so $\delta_{\textrm{run}}(X_{1:n}) \rightarrow \infty$. Thus the critic
in Proposition~\ref{prop:critic_longest_run} captures atypicality associated with the longest run.
As a more subtle example, suppose that $X_{1:n}$ is generated according to the
distribution in which bits are selected independently and uniformly except that 
if a run of length $\log \log n$ of either zeros or ones occurs, then the next
symbol deterministically ends the run.\footnote{This process is a reasonable
model for human-generated sequences of random bits~\cite{SchillingRunCollegeMath,Revesz1978StrongTheoremsICM}.} For any $k$, the $k$-th order empirical
distribution of this process will eventually be close to that of $k$ i.i.d.\ 
uniform bits. But this distribution's deviation from true Bernoulli trials
will eventually be detected by $\delta_{\textrm{run}}$.

Critics can also be obtained from lossless compressors. Let $f : \mathcal{X} \rightarrow \{0,1\}^*$
denote a prefix-free lossless compression algorithm, such as one of the various
Lempel-Ziv algorithms~\cite[Sec.~13.4]{Cover&Thomas2006}. Then
\begin{IEEEeqnarray}{c}
\delta(x) = \log\frac{1}{p_X(x)} - |f(x)|,
\label{eq:critic_based_on_compressor}
\end{IEEEeqnarray}
where $|f(x)|$ is the length of $f(x)$, is a $p_X$-critic by
Kraft's inequality~\cite[Theorem~5.2.1]{Cover&Thomas2006}. This critic measures the extent to which
$x$ is atypically difficult to compress under $f$.
Measures of this form have been shown to be relevant for deep learning based outlier detection \cite[Section~5.1]{2024TheisUniversalCriticsPositionPaper}.
Replacing $|f(x)|$ in \eqref{eq:critic_based_on_compressor} by the Kolmogorov complexity $K(x)$ would also give a valid $p_X$-critic\footnote{Here, the resulting critic is defined over the finite set $\mathcal{X}.$ The expression for the universal critic mentioned in the Section \ref{sec:intro} is substantially similar, but involves a countably infinite set, which is essential
to the theoretical notion of universality.
The details are provided in
Appendices \ref{app:subsec_universal_critics_and_semi_measures}
and
\ref{app:proof_of_existence_universal_critic}
--- see also
\cite[Section~4]{2024TheisUniversalCriticsPositionPaper} for a high-level and insightful presentation.
}
\cite[Lemma~14.3.1]{Cover&Thomas2006}.

Evidently the set of computable critics is quite expansive. Yet it is
worth noting that it does not include every critic of interest.
Fix a sequence of codes
(formally defined in Section \ref{sec:formulation_asymptotic_achievability} to follow)
and let $q_n$ denote the distribution
of the reconstruction when the blocklength is $n$.
Define
a
$p^{\otimes *}$-critic
as
$
\delta_\star(x_{1:n}) = \log
q_n(x_{1:n})
/
p_X^{\otimes n}(x_{1:n})
,
$
for $x_{1:n}$ in $\mathcal{X}^n$.
This is a critic that directly tests whether
$x_{1:n}$ is more likely to be a source string or a reproduction.  This 
critic is not computable in general because it requires 
implementing a countable family of algorithms, one for each blocklength. It is
computable if
the sequence of codes is such that $\{q_n\}_n$ is computable, and 
in general, $\delta_\star$ can also
be approximated by a computable critic if we consider very large
batches of sequences --- as formalized in Section \ref{sec:formulation_asymptotic_achievability} and Theorem \ref{thm:very_large_batch_size}, and its proof.
\begin{remark}\label{rem:our_def_of_critic_differs_from_classical_definitions}
Definition \ref{def:sum_critic_at_beginning_of_paper}
is a composite of two prominent definitions
of a $p^{\otimes *}$-critic:
Definitions~4.3.8 and 2.4.1
in
\cite{BookKolmogorovComplexity} --- which is a standard algorithmic information theory textbook.
The former definition is precisely \eqref{eq:conventional_sum_test_requirement}, where weights $\{\pi(n)\}_{n\in\mathbb{N}}$ are assumed to be given.
The second definition (\cite[Definition~2.4.1]{BookKolmogorovComplexity}) is:
\begin{IEEEeqnarray}{c}
\forall n \in \mathbb{N}, \forall C \in \mathbb{N}_{\geq 0}, \
p^{\otimes n}\big(\delta(X) \geq C
\big) \leq 2^{-C},
\nonumber
\end{IEEEeqnarray}
where the $p^{\otimes *}$-critic $\delta$ is required to take only non-negative integer values.
The latter definition is more suitable for a Shannon-theoretic coding formulation, where each blocklength is viewed as a problem unto itself, while the former definition is 
more succinct and more convenient in certain respects
\cite[p.~220]{BookKolmogorovComplexity}.
Our proposed formulation
(Definition~\ref{def:sum_critic_at_beginning_of_paper})
combines the blocklength-independence
of~\cite[Definition~2.4.1]{BookKolmogorovComplexity} with the mathematical convenience of~\cite[Definition~4.3.8]{BookKolmogorovComplexity}.
As such, replacing Definition \ref{def:sum_critic_at_beginning_of_paper} with \cite[Definition~2.4.1]{BookKolmogorovComplexity} would not change the mathematical notion of realism which we adopt in Section \ref{sec:formulation_asymptotic_achievability}, and thus, not change our coding theorems (Theorems \ref{thm:small_batch_size_asymptotics} and \ref{thm:very_large_batch_size} to follow).
\end{remark}

\section{New model for the rate-distortion-perception trade-off}\label{sec:new_RDP_formalism_and_all_results}

\subsection{The one-shot setting}\label{subsec:one_shot_definitions}
\begin{figure}[t!]
\centering\includegraphics[width=0.65\textwidth]{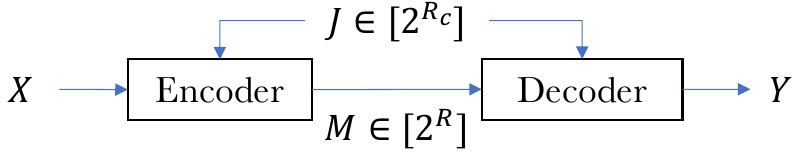}
\caption{The system model for the one-shot setting.}
\label{fig:general_setup_one_shot_compression}
\end{figure}
We
consider a function $d:\mathcal{X}{\times}\mathcal{X} \to [0,\infty)$ called the distortion measure.
We assume that
a compression scheme can use unlimited private randomness at the encoder or decoder,
and also potentially leverage available common randomness $J$ between the encoder and
decoder, as depicted in Figure \ref{fig:general_setup_one_shot_compression}, and formalized in the following definition.
\begin{definition}
Given
non-negative reals $R$ and $R_c,$ an $(R,R_c)$ code is a privately randomized encoder and decoder couple $(F,G)$ consisting of a conditional distribution $F_{M|X, J}$ from $ \mathcal{X} \times [2^{R_c}]$ to $  [2^{R}],$ and a conditional distribution $G_{Y| M,J}$ from $[2^{R}] \times [2^{R_c}]$ to $\mathcal{X}.$
Variables $M$ and $Y$ are called the \textit{message} and \textit{reconstruction}, respectively, and
distribution
$
P 
:=
p_{X} \cdot p^{\mathcal{U}}_{[2^{R_c}]} \cdot 
F_{M|X, J} \cdot G_{Y|M,J}
$
Such a code is said to be deterministic if $R_c=0$ and mappings $F,G$ are deterministic.
\end{definition}

We propose a
new
RDP trade-off,
formalized in the following two definitions.
\begin{definition}\label{def:additive_distortion}
We extend $d$ into an \textit{additive distortion measure} on batches of elements of $\mathcal{X}$:
$$\forall B{\in}\mathbb{N}, \
\forall (\mathbf{x}^{(B)},\mathbf{y}^{(B)}) \in \mathcal{X}^B \times \mathcal{X}^B, \quad d(\mathbf{x}^{(B)}, \mathbf{y}^{(B)}) := \tfrac{1}{B}\scalebox{1.05}{$\sum_{k{=}1}^B$} d(x^{(k)},y^{(k)}).$$ 
\end{definition}
\begin{definition}\label{def:achievability_one_shot_average_realism}
Consider
a positive integer $B,$
and
a
$p_X^{\otimes B}$-critic
$\delta.$
A tuple $(R,
\Delta, C
)$ is said to be $\delta
$-achievable
if
there exists
$R_c\in\mathbb{R}_{\geq 0}$
and
an $(R,
R_c)$ code
such that the induced distribution $P$
satisfies
\begin{IEEEeqnarray}{rCl}
\mathbb{E}_{P^{\otimes B}}
\big[
d(
\mathbf{X}
^{
(
B
)
}
,
\mathbf{Y}
^{
(
B
)
}
)
\big]
\leq
\Delta
\ \text{ and }
\mathbb{E}_{P^{\otimes B}}[
\delta
(
\mathbf{Y}
^{
(
B
)
}
)
]
\leq
C
,
\nonumber
\end{IEEEeqnarray}
where $\mathbf{X}^{(B)}\in\mathcal{X}^B$ denotes a batch of $B$ \iid source samples, and $\mathbf{Y}^{(B)}$ the batch of corresponding reconstructions produced by the code (with each source sample being compressed separately).
If the code is deterministic, then we say that $(R,\Delta,C)$ is $\delta
$-achievable with a deterministic code.
\end{definition}

The main 
difference with the original RDP trade-off of \cite{2019BlauMichaeliRethinkingLossyCompressionTheRDPTradeoff} pertains to the realism constraint. In the
formulation
of \cite{2019BlauMichaeliRethinkingLossyCompressionTheRDPTradeoff}, the realism constraint is $\mathcal{D}(p_X, P_Y) \leq C,$ where 
$\mathcal{D}$ is some divergence.
Intuitively, that constraint 
corresponds to the special case of infinite batch size in the RDP trade-off proposed in Definition \ref{def:achievability_one_shot_average_realism}, since the discrete distributions 
$p_X$ and $P_Y$ can be approximated arbitrarily well using a large enough number of samples.
In that sense, our proposed RDP framework generalizes the original one, 
through involving elements of practical realism metrics, such as the number $B$ of 
samples that are inspected, and a scoring function $\delta$ which is required to be 
approximable via an algorithm.
Theorem \ref{thm:very_large_batch_size} to follow constitutes a rigorous statement of this intuition.
In the next section, we define an asymptotic notion of achievability.
We provide achievable points in the sense of Definition \ref{def:achievability_one_shot_average_realism} in Section \ref{subsec:one_shot_result}.

\subsection{Asymptotic
setting}\label{sec:formulation_asymptotic_achievability}

\begin{figure}[t!]
\centering\includegraphics[width=0.6\columnwidth]{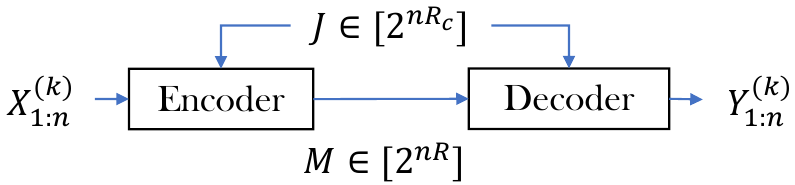}
\caption{The system model for the asymptotic setting. Index $k$ ranges from $1$ to the batch size. The same encoder-decoder pair is used to process each source sample in the batch.}
\label{fig:general_setup_asymptotic_setting}
\end{figure}

We consider the compression of a source distributed according to $p_X^{\otimes n}.$
We characterize the RDP trade-off in asymptotic settings where both $n$ and the batch size go to infinity.
The extension of
$d$ into an additive distortion measure on finite sequences, and batches of finite sequences,
follows from Definition \ref{def:additive_distortion}.
The setup is depicted
in Figure \ref{fig:general_setup_asymptotic_setting}.
A coding scheme is designed for a specific blocklength $n,$ and
a batch
of samples
is
considered when assessing the scheme's performance, as follows.
Given a coding scheme, each item in a batch of source samples is compressed separately, and realism is measured based on the resulting batch of reconstructions. This is formalized in the definitions below.
\begin{definition}\label{def:n_letter_code}
Given
$R,R_c \geq 0,$ and $n\in\mathbb{N},$
an $(n, R, R_c)$ code is a privately randomized encoder and decoder couple $(F^{(n)},G^{(n)})$ consisting of a mapping $F^{(n)}_{M|X_{1:n}, J}$ from $\mathcal{X}^n \times [2^{nR_c}] $ to $ [2^{nR}]$ and a mapping $G^{(n)}_{Y_{1:n}| M,J}$ from $[2^{nR_c}] \times [2^{nR}]$ to $ \mathcal{X}^n.$
Such a code is said to
be fully deterministic if $R_c=0$ and both $F^{(n)}$ and $G^{(n)}$ are deterministic.
The distribution induced by the code is
\begin{IEEEeqnarray}{c}
P^{(n)}
:=
p_{X}^{\otimes n} \cdot p^{\mathcal{U}}_{[2^{nR_c}]} \cdot 
F^{(n)}_{M|X_{1:n}, J} \cdot G^{(n)}_{Y_{1:n}|M,J},
\nonumber
\end{IEEEeqnarray}
and variable $Y_{1:n}$ is called the \textit{reconstruction}.
\end{definition}

We define asymptotic achievability as follows. See Appendix \ref{app:algorithmic_information_theory} for background on
computability.
\begin{definition}
\label{def:achievability_batch_average_statistical_realism_for_all_critics}
A quadruplet $(R, R_c, \{B_n\}_{n \geq 1},
\Delta)$ is said to be asymptotically achievable with \textit{algorithmic realism} if
for any $\varepsilon>0,$
there exists
a sequence of codes $\{(F^{(n)},G^{(n)})\}_n,$ the $n$-th being $(n, R
+\varepsilon, 
R_c),$
such that the induced sequence $\{P^{(n)}\}_n$ of distributions
satisfy
\begin{IEEEeqnarray}{c}
\limsup_{n \to \infty}
\mathbb{E}_{(P^{(n)})^{\otimes 
B_n
}}
\big[
d(
\mathbf{X}^{(n,
B_n
)}
,
\mathbf{Y}^{(n,
B_n
)}
)
\big]
\leq \Delta
+\varepsilon
\nonumber
\label{eq:def_batch_distributional_realsim_distortion_constraint}
\end{IEEEeqnarray}
(where
$
(
\mathbf{X}^{(n,
B_n
)}
,
\mathbf{Y}^{(n,
B_n
)}
)
$
is a batch $\{(X^{(k)}_{1{:}n},Y^{(k)}_{1{:}n})\}_{k {\in} [B_n]}$
of \iid pairs, each distributed according to $P^{(n)}$),
and
such that
for any lower semi-computable $p_X^{\otimes *}$-critic $\delta
,$
\begin{IEEEeqnarray}{c}
\sup_{n\in\mathbb{N}}
\mathbb{E}_{
(P^{(n)})^{\otimes 
B_n 
}
}
\big[
\delta
(\mathbf{Y}^{(n,
B_n
)})
\big]
<\infty
.
\label{eq:def_distributional_realsim_perception_constraint_general_batch_size}
\IEEEeqnarraynumspace
\end{IEEEeqnarray}
We say that $(R,\{B_n\}_{n \geq 1}, \Delta)$ is achievable with \textit{a fully deterministic scheme} if for each $n,$ the code $(F^{(n)},G^{(n)})$ is fully deterministic.
\end{definition}

Note that in the above definitions each source sample $X_{1:n}^{(k)}$ is compressed based on a different realization of the common randomness.
Since a $p$-critic is a log-likelihood ratio in the sense of
\eqref{eq:sum_critic_is_a_likelihood_ratio},
then,
by Jensen's inequality,
for a batch of source samples,
the expected score given by a $p_X^{\otimes *}$-critic $\delta$ satisfies
\begin{IEEEeqnarray}{c}
\forall n\in\mathbb{N}, \
\mathbb{E}_{
(P^{(n)})^{\otimes 
B_n 
}
}
\big[
\delta
(\mathbf{X}^{(n,
B_n
)})
\big]
\le 0.
\nonumber
\end{IEEEeqnarray}
In the examples in the previous section, we observed that when
the distribution of
$\mathbf{Y}^{(n,B_n)}$
lacks certain properties of the source distribution,
then we
expect $\delta(\mathbf{Y}^{(n,B_n)})$ to grow unboundedly with $n$
--- see also standard algorithmic information theory literature, e.g., \cite[Definition~2.4.3]{BookKolmogorovComplexity}.
This 
motivates the choice of requiring the expected realism discrepancy
to
remain bounded with $n$, as opposed to requiring it
to
remain below
some fixed threshold as in Def.~\ref{def:achievability_one_shot_average_realism}.
It is unclear \emph{a priori} that even this
nonspecific constraint can be satisfied at any rate below the entropy
of the source since it must hold for every computable critic. We shall
see
that one can satisfy this constraint with deterministic
codes that achieve the RDP
function. Crucial
to this result is that the bound in \eqref{eq:def_distributional_realsim_perception_constraint_general_batch_size}
is not imposed uniformly over the critics.

\section{Results}\label{sec:statements_of_3_main_theorems}

\subsection{Low batch size regime}\label{subsec:low_batch_size_thm}

The following theorem states that
$R^{(1)}(\cdot),$ defined in \eqref{eq:def_RDP_function}, which naturaly arises in the distribution matching formalism, also characterizes
the optimal trade-off in our asymptotic setting, when the batch size is not impractically large.
\begin{theorem}\label{thm:small_batch_size_asymptotics}
Consider a finite source alphabet $\mathcal{X}$ such that $|\mathcal{X}|\geq 2,$ a computable source distribution $p_X$ on $\mathcal{X}$ such that $\forall x \in \mathcal{X}, p_X(x)>0,$
and
a sequence $\{B_n\}_{n \geq 1}$ of positive integers such that
$
\log(B_n)/n \ \
\substack{\raisebox{-4pt}{$\longrightarrow$} \\ \scalebox{0.7}{$n$$\to$$\infty$}} \ \
0.
$
For any $\Delta\in\mathbb{R}_+,$ let $R(\Delta)$ be the infimum of rates $R$ such that there exists $R_c\in\mathbb{R}_{\geq 0}$ such that $(R, R_c, \{B_n\}_{n \geq 1},
\Delta)$ is asymptotically achievable with \textit{algorithmic realism}.
Moreover,
let $R_*(\Delta)$ be the infimum of rates $R$ such that $(R, \{B_n\}_{n \geq 1},
\Delta)$ is asymptotically achievable with \textit{algorithmic realism} with fully deterministic codes.
Then,
if
$R^{(1)}(\Delta)<H_p(X),$
we have
$
R(\Delta) = R_*(\Delta) = R^{(1)}(\Delta).
$
\end{theorem}

The proof is provided in Appendices \ref{app:achievability_small_batch_size} and \ref{app:converse}.
The strength of this result lies in how stringent constraint \eqref{eq:def_distributional_realsim_perception_constraint_general_batch_size} is: a single compression scheme satisfies a performance guarantee for every
relevant
$p_X^{\otimes *}$-critic,
and deterministic schemes are sufficient.
Moreover, one can find such a scheme for any batch size sequence which is sub-exponential in the dimension $n$ of the source, i.e., for all regimes where the batch size is not impractically large.
To prove the achievability direction of Theorem \ref{thm:small_batch_size_asymptotics}, we leverage the existence of a \textit{universal} $p_X^{\otimes *}$-critic $\delta_0$ (see Appendix \ref{app:subsec_universal_critics_and_semi_measures}), which is one of the great successes of algorithmic information theory. Indeed, it is sufficient to construct a scheme which achieves \eqref{eq:def_distributional_realsim_perception_constraint_general_batch_size} only for such a $\delta_0,$ which is more sensitive than all relevant $p_X^{\otimes *}$-critics.
It is a very strong critic, stronger than can be implemented in practice, which is another strength of Theorem \ref{thm:small_batch_size_asymptotics}.

\subsection{One-shot achievable points}\label{subsec:one_shot_result}

For theoretical interest, we provide
a
family of
points
which are achievable,
in the sense of Definition \ref{def:achievability_one_shot_average_realism},
without any statistical assumption on the source distribution $p_X.$
For the sake of
intuition, one can consider the following example:
\begin{itemize}
    \item $\mathcal{X}$ is a finite set of images, e.g., the set of all images of a given resolution, with a finite range for pixels (finite precision).
    \item $d$ is the mean squared error between pixel values.
    \item $B$ is the number of images inspected by the critic at a time.
    \item $R$ is the number of bits into which a given image is compressed.
\end{itemize}
\begin{theorem}\label{thm:one_shot_achievable_points}
Consider a finite set $\mathcal{X}$ with $|\mathcal{X}|\geq 2,$ a computable distribution $p_X$ on $\mathcal{X}$ such that $\forall x \in \mathcal{X}, p_X(x)>0,$ some $B\in\mathbb{N},$ some
$
(
R
,
\Delta
)
\in\mathbb{R}_+
^2
,$ and a $p_X^{\otimes B}$-critic $\delta.$ Consider any conditional transition kernel $p_{Y|X}$ from $\mathcal{X}$ to $\mathcal{X}$ satisfying
$
p_Y \equiv p_X
$
and
$
\mathbb{E}_p[d(X,Y)] \leq \Delta.
$
Then, for any $\varepsilon \in (0,\Delta/2),$ and any $\gamma>0,$ the triplet $(R,\Delta',C)$ is $\delta$-achievable with an $(R,0)$ code having a deterministic decoder, with
\begin{IEEEeqnarray}{rCL}
\Delta' &:=& \Delta + \varepsilon + \dfrac{6\Delta}{\varepsilon}\max(d) \cdot \eta_{R,\gamma}
\nonumber\\
C &:=&
\scalebox{0.97}{$
\dfrac{3\Delta}{\varepsilon}
\Bigg[
1+
\Big(
\dfrac{B^2}{\lfloor 2^{R} \rfloor}
+ 2B\eta_{R,\gamma}
\Big)
\cdot \max_x B\log\dfrac{2}{p_X(x)}
\Bigg]
$}
\nonumber
\\
\eta_{R,\gamma} &:=& p(\mathcal{A}_{R,\gamma})+2^{-\gamma\log(|\mathcal{X}|)/2}
\nonumber\\
\mathcal{A}_{R,\gamma} &:=& \Big\{(x,y)\in\mathcal{X}^2 \ | \  \log\Big(\dfrac{p_{X,Y}(x,y)}{p_X(x)p_Y(y)}\Big)-\log(\lfloor2^{R}\rfloor) > -\gamma \log(|\mathcal{X}|) \Big\},
\nonumber
\end{IEEEeqnarray}
where $\max(d)$ stand for $\max_{x,y\in\mathcal{X}}d(x,y),$ and we use
the convention $0/0:=1.$
\end{theorem}

The proof is provided in Appendix \ref{app:proof_achievability_one_shot_achievable_points}.
The term $B^2/\lfloor 2^{R} \rfloor$ is an upper bound on the probability that two source samples in the batch are compressed into the same message.
This is related to the so-called \textit{birthday paradox} (see
Claim \ref{claim:birthday_bound_one_shot} in Appendix \ref{app:proof_achievability_one_shot_achievable_points}).
The term $
\max_x B\log(1/p_X(x))$
is an upper bound on the output of $\delta,$ which follows from Definition \ref{def:sum_critic_at_beginning_of_paper}.
Theorem \ref{thm:one_shot_achievable_points}
provides insights on the asymptotic regime of Theorem \ref{thm:small_batch_size_asymptotics}.
Consider
the limit of large $|\mathcal{X}|,$ with fixed $\Delta,\varepsilon,\gamma,$
and $R$ proportional to $\log|\mathcal{X}|,$
and with
$\log(B) = o(\log|\mathcal{X}|).$
We know that
$
\mathbb{E}_p
\big[
\log
p_{X,Y}(x,y)
/
p_X(x)p_Y(y)
\big]
= I_p(X;Y).
$
Hence, if this log-likelihood ratio concentrates well, and if $R > I_p(X;Y),$ as in the definition of $R^{(1)}(\cdot)$ in \eqref{eq:def_RDP_function}, then $p(\mathcal{A}_{R,\gamma})$ is small for small enough $\gamma.$ In such an asymptotic regime,
we obtain $\Delta' \approx \Delta,$ and $C = O(
1
).$
Therefore,
the assumption in Theorem \ref{thm:small_batch_size_asymptotics}, that the source is of the form $p_X^{\otimes n}$ for some large $n,$ is only used to ensure fast concentration of the log-likelihood ratio.
Hence, Theorem \ref{thm:small_batch_size_asymptotics} can be extended
to a larger set of sources.
In the next section, we present our last main result, which pertains to an asymptotic regime with large batch size.

\subsection{Generalizing the distribution matching formalism}

We present a result which connects our proposed formalism for the RDP trade-off to the distribution matching formalism of \cite{2019BlauMichaeliRethinkingLossyCompressionTheRDPTradeoff}, and concludes our findings regarding the role of common randomness.

\vspace{5pt}
\subsubsection{Background}\label{sec:short_background_on_distribution_matching}
Under the
distribution matching formalism for the RDP trade-off,
the natural asymptotic notion of achievability is as follows.
\begin{definition}\label{def:near_perfect_realism}(\cite{Sep2015RDPLimitedCommonRandomnessSaldi,2019BlauMichaeliRethinkingLossyCompressionTheRDPTradeoff})
A triplet $(R, R_c,
\Delta)$ is said to be asymptotically achievable with \textit{near-perfect realism} if
for any $\varepsilon>0,$
there exists
a sequence of codes $\{(F^{(n)},G^{(n)})\}_n,$ the $n$-th being $(n, R
+\varepsilon, 
R_c),$
such that the sequence $\{P^{(n)}\}_n$ of distributions induced by the codes satisfies
\begin{IEEEeqnarray}{c}
\limsup_{n \to \infty}
\mathbb{E}_{
P^{(n)}
}
\big[
d(
X_{1:n}
,
Y_{1:n}
)
\big]
\leq \Delta
+\varepsilon,
\nonumber
\\*
\scalebox{0.99}{$
\|P^{(n)}_{Y_{1:n}} - p_{X}^{\otimes n}\|_{TV} \underset{n \to \infty}{\longrightarrow} 0.
$}
\label{eq:def_near_perfect_realism_constraint}
\IEEEeqnarraynumspace
\end{IEEEeqnarray}
\end{definition}

TVD in \eqref{eq:def_near_perfect_realism_constraint} is directly related to the performance of the optimal hypothesis tester between the reconstruction distribution $P^{(n)}_{Y_{1:n}},$ and the source distribution $p_X^{\otimes n}$ \cite{2019BlauMichaeliRethinkingLossyCompressionTheRDPTradeoff}.
Replacing \eqref{eq:def_near_perfect_realism_constraint} with
\begin{IEEEeqnarray}{c}
\exists N \in \mathbb{N}, \forall n \geq N, \ P^{(n)}_{Y_{1:n}} \equiv p_X^{\otimes n}
\nonumber
\end{IEEEeqnarray}
gives the notion of asymptotic \textit{achievability with perfect realism}.
These two notions are equivalent for finite-valued sources \cite{Sep2015RDPLimitedCommonRandomnessSaldi}, as well as for continuous sources under mild assumptions \cite{Sep2015RDPLimitedCommonRandomnessSaldi,2022AaronWagnerRDPTradeoffTheRoleOfCommonRandomness}.

\vspace{5pt}
\subsubsection{Connection to our formalism}
As
stated
in the theorem below, in a certain large batch size regime, asymptotic achievability with algorithmic realism (Definition \ref{def:achievability_batch_average_statistical_realism_for_all_critics}) is equivalent to asymptotic achievability with near-perfect realism (Definition \ref{def:near_perfect_realism}).
The proof is provided in Appendix \ref{app:proof_thm_large_batch_size_asymptotics}.
\begin{theorem}\label{thm:very_large_batch_size}
Consider
a computable
increasing
sequence $\{B_n\}_{n \geq 1}
$ of positive integers such that
$
B_n
/
|\mathcal{X}|^n
\
\substack{\raisebox{-4pt}{$\longrightarrow$} \\ \scalebox{0.7}{$n$$\to$$\infty$}} \
\infty.
$
Then, for any $R_c \in \mathbb{R}_{\geq 0},$ and any $(R,\Delta)\in(\mathbb{R}_+)^2,$ tuple $(R, R_c, \{B_n\}_{n \geq 1},
\Delta)$ is asymptotically achievable with \textit{algorithmic realism} if and only if $(R,R_c,\Delta)$ is asymptotically achievable with near-perfect realism, if and only if $(R,R_c,\Delta)$ is asymptotically achievable with perfect realism.
The closure of the set of such achievable triplets  $(R,R_c,\Delta)$ is equal to the following region:
\begin{align*}
      & 
    \left\{ \begin{array}{rcl}
        \scalebox{1.0}{$\exists \  p_{Y,V|X}$} &:& \scalebox{1.0}{$p_{Y} \equiv p_X$}\\
        \scalebox{1.0}{$R$} &\geq& \scalebox{1.0}{$I_p(X;V)$} \\
        \scalebox{1.0}{$R+R_c$} &\geq& \scalebox{1.0}{$I_p(Y;V)$} \\
        \scalebox{1.0}{$\Delta$} &\geq& \scalebox{1.0}{$\mathbb{E}_p[d(X, Y)]$}\\
        X \quad - &V& - \quad Y
    \end{array}\right\}.
\end{align*}
\end{theorem}

Hence, Theorem \ref{thm:very_large_batch_size}, similarly to the finding in \cite{2024TheisUniversalCriticsPositionPaper}, shows that for large batch size,
our
formalism
is equivalent to the
distribution matching
formalism.
Hence, the former is a generalization of the latter.

\section{Conclusion}
\label{sec:conclusion}

In this work, we have proposed a theoretical framework for realism constraints in lossy compression that is more in line with human perception than the distribution matching formalism.
Indeed, our formulation incorporates key aspects of practical realism metrics: realization-based critics that can be algorithmically computed, and the number (batch size) of samples jointly inspected by a critic.
Inspired by the algorithmic information theory literature and \cite{2024TheisUniversalCriticsPositionPaper}, we require that reconstructions asymptotically satisfy all computable critics of randomness. This includes critics inspecting the frequency of any pattern, but also critics inspecting subtle properties, such as ones pertaining to the Erd\H{o}s-R\'{e}nyi law of runs.
Theorem \ref{thm:small_batch_size_asymptotics} states that common randomness does not improve the trade-off under our formalism, in all regimes where the batch size is not impractically large with respect to the dimension $n$ of the source.
This offers a compelling explanation for
the fact that the need for large amounts of common randomness has not been observed in practice.
Theorem \ref{thm:very_large_batch_size} states that
when the batch size is extremely large,
our formalism is equivalent to the distribution matching formalism \cite{Sep2015RDPLimitedCommonRandomnessSaldi,2022AaronWagnerRDPTradeoffTheRoleOfCommonRandomness,Dec2022WeakAndStrongPerceptionConstraintsAndRandomness},
hence
common randomness is useful in that regime.
Thus, Theorems \ref{thm:small_batch_size_asymptotics} and \ref{thm:very_large_batch_size}
indicate that, in order to understand the role of randomization in lossy compression with realism constraints,
the focus should be shifted
to the size of the batch inspected by the critic.
A continuation of our work could be to investigate
realism metrics, with particular attention given to the choice of the batch size. This could lead to highlighting specific strengths and weaknesses of existing realism metrics. It may also inspire a critical assessment of the relative performance of existing compression schemes, depending on the choice of the realism metric. Another future direction could be to more precisely characterize the amount of randomness needed as a function of the batch size.
Furthermore,
possible extensions of our setup include compression with side information \cite{2023YassineGunduzISITRDPSideInformation}, and other distributed scenarios.

\appendices

\section{Further background on algorithmic information theory}\label{app:algorithmic_information_theory}

\subsection{Computability}

The following
definition matches \cite[Definition~1.7.4]{BookKolmogorovComplexity}, except for the definition of a computable real number, which we adapted from \cite[Exercise~1.7.22]{BookKolmogorovComplexity}, and the definition of a computable set, which matches that of \cite[page 32]{BookKolmogorovComplexity}.
\begin{definition}\label{def:semi_computability}
Common countable sets, such as $\mathbb{N}_{\geq 0}^k$ for $k\in\mathbb{N},$ and $\{0,1\}^*,$ and subsets thereof, are implicitly identified with
$\mathbb{N}_{\geq 0}$
and subsets thereof
via some reference bijections.
Thereby, the notions formalized in the present definition
extend to functions having such sets as domain or range.
Consider a subset $\mathcal{E}$ of $\mathbb{N}_{\geq 0}.$
A map $f$ from
$\mathcal{E}$
into
$\mathbb{N}_{\geq 0}^3$
is said to be \textit{computable} if
there exists
a Turing machine
\cite[Section~1.7.1]{BookKolmogorovComplexity}
such that for any $a\in\mathcal{E},$ the machine halts and outputs $f(a).$
Consider
a computable map $f$ from
a subset
$\mathcal{E}$
of $\mathbb{N}_{\geq 0}$
into
$\{0,1\}\times\mathbb{N}_{\geq 0} \times \mathbb{N}.$
Then,
composing
$f$
with $(s,a,b) \mapsto (2s-1)a/b \ $
yields
a map from $\mathcal{E}
$ to $\mathbb{Q},$ which
is said to be a \textit{computable map from} $\mathcal{E}
$ \textit{to} $\mathbb{Q}.$
A map $f$ from a subset $\mathcal{E}$ of $\mathbb{N}_{\geq 0}$ into $\mathbb{R}$ is said to be \textit{lower semi-computable} if
there exists a
computable function
$\varphi$
from $\mathcal{E}\times\mathbb{N}$ into $\mathbb{Q},$ such that
$$\forall x \in \mathcal{E}, \varphi(x,k) \underset{k\to\infty}{\rightarrow}f(x), \qquad \text{and} \qquad \forall x \in \mathcal{E},\forall k \in \mathbb{N}, \ \ \varphi(x,k+1) \geq \varphi(x,k).$$
Moreover, $f$ is said to be a \textit{computable map from} $\mathcal{E}
$ \textit{to} $\mathbb{R}$ if both $f$ and $-f$ are lower semi-computable.
A real number $\lambda$ is said to be computable if the constant function $f:\mathbb{N}_{\geq 0}\to\mathbb{R}, \ n \mapsto \lambda$ is a computable function from $\mathbb{N}_{\geq 0}$ to $\mathbb{R}.$
A (possibly infinite) subset $\mathcal{X}$ of $\mathbb{N}_{\geq 0},$ is said to be computable if there exists a computable function $f$ from $\mathbb{N}_{\geq 0}$ to $\{0,1\},$ which returns $1$ if its input is in $\mathcal{X},$ and $0$ otherwise.
\end{definition}

The following lemma allows us to construct (semi-)computable functions.
A partial proof is provided in
Appendix \ref{app:proof_practical_semicomputability_lemma}.
\begin{lemma}\label{lemma:semicomputability_preserved_by_certain_operations}
Let $\mathcal{E}$ denote a non-empty subset of $\mathbb{N}_{\geq 0},$ and let $f$ and $g$ denote functions from $\mathcal{E}$ to $\mathbb{R}.$\\
(i) If $f$ and $g$ are both lower semi-computable, then functions $f+g,$ $\lceil f \rceil,$ and $2^f$ are lower semi-computable.
If, in addition, $f$ and $g$ only take non-negative values, then $fg$ and $2^f/(3+f)^2$ are lower semi-computable. If, in addition, $f$ only takes positive values, then $\log(f)$ is lower semi-computable.\\
(ii) If $f$ and $g$ are both computable, then
$f+g,$ $fg,$ and $|f|$ are computable. If, in addition, $f$ only takes positive values, then functions $1/f,$ $\log(f),$ and $f^{1/b}$ are computable, for any positive integer $b.$\\
(iii) Let $\mathcal{X}$ be a
finite subset of $\{0,1\}^s$ for some positive integer $s.$ If $f
:
\{0,1\}^*
\to
\mathbb{R}
$ is
lower semi-computable,
then
function $\Tilde{f}: \{0,1\}^* \to \mathbb{R}$ which is null outside of $\cup_{n \in \mathbb{N}}\mathcal{X}^n,$ and defined by $$\forall x \in \cup_{n \in \mathbb{N}}\mathcal{X}^n, \ \ \Tilde{f}(x) = \sum_{y \in \mathcal{X}^{l(x)}} f(y),$$ is lower semi-computable.
Moreover, if $p$ is a lower semi-computable probability measure on $\mathcal{X},$ then $p^{\otimes *}$ is lower semi-computable.
\end{lemma}

\subsection{Universal critics and semi-measures}\label{app:subsec_universal_critics_and_semi_measures}

\begin{definition}
Given a finite set $\mathcal{W},$ a function $f:\mathcal{W}\to[0,1]$ is a \textit{semi-measure} if $$\sum_{w\in\mathcal{W}} f(w) \leq 1.$$ It is said to be a \textit{lower semi-computable semi-measure} if $f$ is a semi-measure and
lower semi-computable.
\end{definition}

The following theorem,
corresponds to Definition 4.3.2, Equation (4.2), and Theorems 4.3.1 and 4.3.3 in \cite{BookKolmogorovComplexity} --- which is a standard algorithmic information theory textbook.
It introduces the notion of \textit{universal $p^{\otimes *}$-critic}, used in \cite{2024TheisUniversalCriticsPositionPaper}.
\begin{theorem}\label{thm:universal_semi-measure_and_critic}
Consider
a finite set $\mathcal{X}.$
Let $p$ be a computable distribution on $\mathcal{X}$ such that $\forall x \in \mathcal{X}, p(x)>0.$
Then,
there exists
a $p^{\otimes *}$-critic
$\delta_0$
(which is not necessarily lower semi-computable),
such that
for any lower semi-computable $p^{\otimes *}$-critic $\delta,$ there exists a constant $c_\delta$ such that
\begin{IEEEeqnarray}{c}
\forall x \in \bigcup_{n \in \mathbb{N}} \mathcal{X}^n, \quad \delta_0(x) \geq \delta(x) - c_\delta.
\label{eq:property_defining_universal_critics}
\end{IEEEeqnarray}
Any $p^{\otimes *}$-critic satisfying \eqref{eq:property_defining_universal_critics} is called a \textit{universal $p^{\otimes *}$-critic}.
\end{theorem}

Since our definitions are slightly different from the classical ones --- see Remark \ref{rem:our_def_of_critic_differs_from_classical_definitions} ---, we provide a proof of Theorem \ref{thm:universal_semi-measure_and_critic}
in Appendix \ref{app:proof_of_existence_universal_critic}.
See Remark \ref{rem:comparing_universal_critics_from_different_definitions} for a comparison of $p^{\otimes *}$-critics that are universal according to different formulations.
Such a critic $\delta_0$ is one of the best measures of realism deficiency according to $p,$ in the limit of arbitrarily long strings.
If a critic $\delta$ identifies a certain amount of deficiency in a given string, then $\delta_0$ will identify at least as much deficiency, up to some additive constant.
Intuitively, $\delta_0$ is sensitive to all properties of randomness according to $p.$
The existence of such a $\delta_0$ constitutes a remarkable property of the set of all lower semi-computable $p^{\otimes *}$-critics (which is infinite).
\begin{remark}\label{remark:ties_to_Kolmogorov_complexity}
The proof of Theorem \ref{thm:universal_semi-measure_and_critic} (Appendix \ref{app:proof_of_existence_universal_critic}) involves a mixture $\mathbf{m}$ called a \textit{universal semi-measure}. Such a measure can be used as a prior distribution, which has been shown to be relevant in machine learning applications involving realism, such as outlier detection and generative modeling \cite{2024TheisUniversalCriticsPositionPaper}.
The universal semi-measure $\mathbf{m}$ can be chosen in such a way that
$
\forall x \in \{0,1\}^*, \ |-\log(\mathbf{m}(x)) - K(x)| \leq c,
$
for some constant $c$ \cite[Theorem~4.3.3]{BookKolmogorovComplexity}, where $K$ is the Kolmogorov complexity
\cite[Section~3.1]{BookKolmogorovComplexity}.
This
constitutes a strong result, since the Kolmogorov complexity is only defined up to a bounded term \cite[Theorem~14.2.1]{Cover&Thomas2006}.
For distributions $p$ having non-zero mass at
a countably infinite number of points,
the map
$x \mapsto \log(1/p(x))-K(x)
$
is sometimes considered to be an approximation of a universal
critic,
see
Remarks \ref{rem:comparing_universal_critics_from_different_definitions} and \ref{remark:careful_not_to_forget_normalization} (in Appendix \ref{app:proof_of_existence_universal_critic} to follow) for details. See also
\cite[Section~4]{2024TheisUniversalCriticsPositionPaper} for a high-level and insightful presentation.
\end{remark}


\section{Proof of Theorem \ref{thm:one_shot_achievable_points}}
\label{app:proof_achievability_one_shot_achievable_points}

\subsection{Outline}
To show the achievability of a tuple $(R,\Delta,C),$
we consider a set of random reconstructions, and study its realism properties in Appendix \ref{app:subsec_novel_soft_covering_lemma_one_shot}. Then, we show the existence of a suitable choice of realizations of the latter reconstructions in Appendix \ref{app:subsec_Q_and_distortion_and_soft_covering_one_shot}.
In Appendix \ref{app:subsec_achievability_proof_using_the_existence_of_Q_one_shot}, we prove Theorem \ref{thm:one_shot_achievable_points} by proposing a compression scheme achieving a close-to-uniform sampling from the set of reconstructions.
For the remainder of Appendix \ref{app:proof_achievability_one_shot_achievable_points}, we fix a finite set $\mathcal{X}$ such that $|\mathcal{X}|\geq 2,$ a computable distribution $p_X$ on $\mathcal{X}$ such that $\forall x \in \mathcal{X}, p_X(x)>0,$ a positive integer $B,$ and a $p_X^{\otimes B}$-critic $\delta.$

\subsection{Realism performance of a uniformly sampled batch of random reconstructions}\label{app:subsec_novel_soft_covering_lemma_one_shot}

\subsubsection{Random candidate reconstructions}\label{app:subsubsec_codebook_construction_one_shot}
Given a positive real $R,$
let $\mathcal{C}$ be a
family of
$\lfloor 2^{R} \rfloor$
\iid variables, each
sampled from $p_{X}.$ The $m$-th variable is denoted
$y(\mathcal{C},m).$ 
We denote
their joint
distribution by $Q_{\mathcal{C}}.$
Given a realization $c$ of $\mathcal{C},$ we consider a batch $\mathbf{y}^{(B)}$ of $B$ elements of $c,$ sampled uniformly with replacement. Then, we compute the batch's realism score $\delta(\mathbf{y}^{(B)}).$
This is formalized in the following lemma, which gives an upper bound of the expected score with respect to $Q_{\mathcal{C}}.$
\begin{lemma}\label{lemma:novel_soft_covering_one_shot}
Consider a positive real $R
,$
and
the following
probability mass function.
\begin{IEEEeqnarray}{c}
Q_{\mathcal{C},\mathbf{M}^{(B)},
\mathbf{Y}^{(B)}}
\: \big(\{y(m')\}_{m'\in[\lfloor 2^{R} \rfloor]},\mathbf{m}^{(B)},
\mathbf{y}^{(B)}\big)
:= \
\Big(
\prod_{m'=1}^{\lfloor 2^{R} \rfloor} 
p_{\scalebox{0.7}{$X$}}
\scalebox{0.8}{$(y(m'))$} 
\Big)\cdot
\scalebox{0.8}{$\dfrac{1}{\lfloor 2^{R} \rfloor^B}
$}
\cdot
\prod_{k=1}^B
\mathbf{1}_{y^{(k)}{=}y(m^{(k)})}.
\nonumber
\end{IEEEeqnarray}
Then,
with
$
\delta_+(\mathbf{Y}^{(B)}) := \max(0, \delta(\mathbf{Y}^{(B)}))
,
$
we have
\begin{IEEEeqnarray}{c}
\scalebox{0.95}{$
\mathbb{E}_{Q}[\delta_+(
\mathbf{Y}^{(B)}
)]
\leq
1+ \dfrac{B^2}{\lfloor 2^{R} \rfloor}
\max_x B  \log\dfrac{2}{p_X(x)}
$}
.
\nonumber
\IEEEeqnarraynumspace
\end{IEEEeqnarray}
\end{lemma}

The remainder of Appendix \ref{app:subsec_novel_soft_covering_lemma_one_shot} is dedicated to the proof of Lemma \ref{lemma:novel_soft_covering_one_shot}. Fix $R>
0
.$

\vspace{5pt}
\subsubsection{Preliminaries}
\begin{claim}\label{claim:upper_bound_of_average_score_uniform_over_all_delta}
For any distribution $p$ on a finite set, any $p$-critic $\delta$ satisfies
\begin{IEEEeqnarray}{c}
\mathbb{E}_p[2^{\delta_+(X)}]\leq 2
\label{eq:positive_part_critic_exponential_moment_bound} \\
\mathbb{E}_p[\delta_+(X)]\leq 1.
\label{eq:positive_part_critic_moment_bound} 
\\
\max(\delta_+) \leq \max_x \log\dfrac{
2}{p_X(x)}
\label{eq:upper_bound_on_delta}.
\end{IEEEeqnarray}
\end{claim}
\begin{IEEEproof}
From \eqref{eq:bound_in_def_sum_critic_if_supported_on_fixed_length_set}, we have
$1 \ge \mathbb{E}_p[2^{\delta(X)}].$
Thus, we have
\begin{IEEEeqnarray}{rCl}
1
\ge
\mathbb{E}_p[2^{\delta(X)}]
\ge
\mathbb{E}_p[2^{\delta_+(X)}\mathbf{1}_{\delta(X) \ge 0}]
& = &
\mathbb{E}_p[2^{\delta_+(X)}\mathbf{1}_{\delta(X) \ge 0} +\mathbf{1}_{\delta(X) < 0} ] - \mathbb{E}_p[\mathbf{1}_{\delta(X) < 0}]
\nonumber\\
& = & \mathbb{E}_p[2^{\delta_+(X)}] - \mathbb{E}_p[\mathbf{1}_{\delta(X) < 0}] \nonumber\\
& \ge &  \mathbb{E}_p[2^{\delta_+(X)}] - 1.\nonumber
\end{IEEEeqnarray}
This establishes \eqref{eq:positive_part_critic_exponential_moment_bound}, which
implies \eqref{eq:positive_part_critic_moment_bound} by Jensen's inequality.
From \eqref{eq:positive_part_critic_exponential_moment_bound}, for any $x$ in $\mathcal{X},$ we have
$2 \ge \mathbb{E}_p[2^{\delta_+(X)}] \geq p(x)2^{\delta_+(x)}.$
This proves \eqref{eq:upper_bound_on_delta}.
\end{IEEEproof}

\vspace{5pt}
\subsubsection{Realism performance}\label{app:subsubsec_batch_realism_performance_of_Q_one_shot}
\begin{claim}\label{claim:birthday_bound_one_shot}
A simple bound yields,
\begin{IEEEeqnarray}{c}
\scalebox{0.95}{$
(p^{\mathcal{U}}_{[\lfloor 2^{R} \rfloor]})^{\otimes 
B
}(M^{(1)},...,M^{(
B
)} \text{ are distinct}) \geq 1 - \dfrac{B^2}{\lfloor 2^{R} \rfloor}.
$}
\nonumber
\end{IEEEeqnarray}
\end{claim}

The proof is straightforward and is omitted.
From the definition
of $Q
,$
for any
Borel set
$\mathcal{E}
\subseteq \mathbb{R}
,$
\begin{IEEEeqnarray}{c}
Q
\Big(\Big\{\delta_0\big(
\big\{y(\mathcal{C},M^{(k)})\big\}_{k \in [
B
]}
\big)
\in \mathcal{E}
\Big\}
\
\Big|
\
\Big\{M^{(1)},...,M^{(
B
)} \text{ are distinct}\Big\}
\Big)
=
p_X^{\otimes
B
}\Big(\delta_0\big(
\mathbf{X}^{(
B
)}
\big)
\in \mathcal{E}
\Big).
\nonumber
\end{IEEEeqnarray}
Therefore, we have
\begin{IEEEeqnarray}{rCl}
\mathbb{E}_{Q}[\delta_{
+}(\{y(\mathcal{C},M^{(k)})\}_{k\in[B]})]
&=&
\sum_{\mathbf{m}^{(B)}} \mathbb{E}_{Q}[\mathbf{1}_{\mathbf{M}^{(B)}{=}\mathbf{m}^{(B)}}\delta_{
+}(\{y(\mathcal{C},m^{(k)})\}_{k\in[B]})]
\nonumber\\*
&=&
\sum_{\mathbf{m}^{(B)}} \mathbb{E}_{Q}[\mathbf{1}_{\mathbf{M}^{(B)}{=}\mathbf{m}^{(B)}}] \mathbb{E}_{Q}[\delta_{
+}(\{y(\mathcal{C},m^{(k)})\}_{k\in[B]})]
\nonumber\\*
&=&
\sum_{\{m^{(k)}\}_{k\in[B]}
\text{ that are }
\neq}
(p^{\mathcal{U}}_{[\lfloor 2^{R} \rfloor]})^{\otimes 
B}(\mathbf{M}^{(B)}{=}\mathbf{m}^{(B)})
\mathbb{E}_{p_X^{\otimes
B
}
}
[\delta_{
+}(\mathbf{X}^{(B)})]
\nonumber\\*
&+&
\sum_{\{m^{(k)}\}_{k\in[B]} \text{ that are not }\neq}
(p^{\mathcal{U}}_{[\lfloor 2^{R} \rfloor]})^{\otimes 
B}(\mathbf{M}^{(B)}{=}\mathbf{m}^{(B)})
\mathbb{E}_{Q}[\delta_{
+}(\{y(\mathcal{C},m^{(k)})\}_{k\in[B]})]
\nonumber\\*
&\leq&
\mathbb{E}_{p_X^{\otimes
B
}
}
[\delta_{
+}(\mathbf{X}^{(B)})]
+
\max(\delta_{
+}) (p^{\mathcal{U}}_{[\lfloor 2^{R} \rfloor]})^{\otimes 
B}
(M^{(1)},...,M^{(
B
)} \text{not }\neq)
\nonumber\\*
&\leq&
{
1} +
\dfrac{B^2}{\lfloor 2^{R} \rfloor}
\max_x B\log\dfrac{
2}{p_X(x)}
,\label{eq:final_bound_on_average_score_over_codebook_using_birthday_paradox}
\end{IEEEeqnarray}
where \eqref{eq:final_bound_on_average_score_over_codebook_using_birthday_paradox} follows from
\eqref{eq:bound_in_def_sum_critic_if_supported_on_fixed_length_set}
and
Claims
\ref{claim:upper_bound_of_average_score_uniform_over_all_delta}
and
\ref{claim:birthday_bound_one_shot}.
This concludes the proof of Lemma \ref{lemma:novel_soft_covering_one_shot}.


\subsection{Further properties of a uniformly sampled batch}\label{app:subsec_Q_and_distortion_and_soft_covering_one_shot}

\begin{proposition}\label{prop:half_of_the_one_shot_proof}
Consider a finite set $\mathcal{X}$
with $|\mathcal{X}|\geq 2,$ a distribution $p_X$ on $\mathcal{X}$ such that $\forall x \in \mathcal{X}, p_X(x)>0,$ some $B\in\mathbb{N},$ some
$
(
R
,
\Delta
)
\in\mathbb{R}_+
^2
,$ and a $p_X^{\otimes B}$-critic $\delta.$ Consider any conditional transition kernel $p_{Y|X}$ from $\mathcal{X}$ to $\mathcal{X}$ satisfying
$
p_Y \equiv p_X
\text{ and }
\mathbb{E}_p[d(X,Y)] \leq \Delta.
$
Then, for any $\varepsilon \in (0,\Delta/2),$ and
$\gamma>0,$
there exists
a family $\{y(m)\}_{m\in[\lfloor 2^{R} \rfloor]},$ denoted $\mathbf{c},$ of elements of $\mathcal{X},$
such that the distribution
\begin{IEEEeqnarray}{c}
Q_{
M,Y,X
}
\: \big(
m,y,x
\big)
:= \
\scalebox{0.8}{$\dfrac{1}{\lfloor 2^{R} \rfloor
}
$}
\cdot
\Big(
\mathbf{1}_{y
{=}y(m
)}
\Big)
\cdot
p_{X|Y{=}y(m
)}(x
)
\quad \text{satisfies}
\IEEEeqnarraynumspace\label{eq:in_main_prop_def_Q_one_shot}
\end{IEEEeqnarray}
\begin{IEEEeqnarray}{rCl}
\|Q_{X
} -
p_{X}
\|_{TV}
&\leq&
\dfrac{3\Delta}{\varepsilon}[p(\mathcal{A}_{R,\gamma})+2^{-\gamma\log(|\mathcal{X}|)/2}]
\label{eq:in_main_prop_TV_Q_X_one_shot}
\\*
\mathbb{E}_{Q^{\otimes B}}[d(\mathbf{X}^{(B)},\mathbf{Y}^{(B)})] 
&\leq&
\Delta + \varepsilon
\label{eq:in_main_prop_distortion_one_shot}
\\*
\mathbb{E}_{Q^{\otimes B}}[\delta(
\mathbf{Y}^{(B)}
)]
&\leq&
\dfrac{3\Delta}{\varepsilon}
\cdot \left( {\
1+}
\dfrac{B^2}{\lfloor 2^{R} \rfloor}
\max_x B\log\dfrac{
2}{p_X(x)}\right),
\label{eq:in_main_prop_realism_one_shot}
\IEEEeqnarraynumspace
\end{IEEEeqnarray}
\begin{IEEEeqnarray}{c}
\text{where} \quad
\mathcal{A}_{R,\gamma}:= \Big\{(x,y)\in\mathcal{X}^2 \ | \  \log\Big(\dfrac{p_{X,Y}(x,y)}{p_X(x)p_Y(y)}\Big)-\log(\lfloor2^{R}\rfloor) > -\gamma \log(|\mathcal{X}|) \Big\}.
\nonumber
\end{IEEEeqnarray}
\end{proposition}
\begin{IEEEproof}
Fix
some $R>
0
,$
$\Delta>0,$
$\varepsilon \in (0,\Delta/2),$
$\gamma>0,$
and
some
$p_{Y|X}$
satisfying
\begin{IEEEeqnarray}{c}
p_Y \equiv p_X, \quad \mathbb{E}_p[d(X,Y)] \leq \Delta.\label{eq:introducing_R_Delta_B_in_proof_main_prop_one_shot}
\end{IEEEeqnarray}
We
apply Lemma \ref{lemma:novel_soft_covering_one_shot}, and use the notation therein.
Then, from Markov's inequality, we have
\begin{IEEEeqnarray}{c}
Q_{\mathcal{C}}
\left(
\mathbb{E}_{Q}[\delta_{
+}(
\mathbf{Y}^{(B)}
)
| \mathcal{C}
]
\geq \dfrac{3\Delta}{\varepsilon}\left({
1+}\dfrac{B^2}{\lfloor 2^{R} \rfloor}
\max_x B\log\dfrac{
2}{p_X(x)}\right)
\right)
\leq
\dfrac{\varepsilon}{3\Delta}
.
\IEEEeqnarraynumspace\label{eq:average_score_concentrates_Markov_ineq_one_shot}
\end{IEEEeqnarray}
We extend the distribution $Q$ as follows.
\begin{IEEEeqnarray}{c}
Q_{\mathcal{C},\mathbf{M}^{(B)},
\mathbf{Y}^{(B)},\mathbf{X}^{(B)}}
\: \big(\{y(m')\}_{m'\in[\lfloor 2^{R} \rfloor]},\mathbf{m}^{(B)},
\mathbf{y}^{(B)},\mathbf{x}^{(B)}\big)
:=
\qquad\qquad\qquad\qquad\qquad\qquad\qquad\qquad\qquad
\nonumber\\*
\qquad\qquad\qquad\qquad\qquad\qquad
Q_{\mathcal{C},\mathbf{M}^{(B)},
\mathbf{Y}^{(B)}}
\: \big(\{y(m')\}_{m'\in[\lfloor 2^{R} \rfloor]},\mathbf{m}^{(B)},
\mathbf{y}^{(B)}\big)
\cdot
\prod_{k{=}1}^B p_{X|Y{=}y(m^{(k)})}(x^{(k)}).
\nonumber
\end{IEEEeqnarray}
Distribution $Q_{\mathcal{C},M^{(1)},Y^{(1)},X^{(1)}}$ corresponds to the setting of \cite[Theorem~VII.1]{2013PaulCuffDistributedChannelSynthesis}, known as the soft covering lemma. 
Since $p_Y \equiv p_X,$ the latter lemma yields that
for any $\tau\in\mathbb{R},$
\begin{IEEEeqnarray}{c}
\mathbb{E}_{\mathcal{C}}\big[\|Q_{X^{(1)}
|\mathcal{C}} - 
p_{X}\|_{TV}\big]
\leq
p(\mathcal{A}_\tau) + 2^{\tau/2}
,
\IEEEeqnarraynumspace\label{eq:expected_TV_Q_X_on_shot}
\end{IEEEeqnarray}
\begin{IEEEeqnarray}{c}
\text{where} \quad
\mathcal{A}_\tau := \{(x,y) \ | \ \log(p_{Y|X{=}x}(y)/p_X(y))-\log(\lfloor2^{R}\rfloor)>\tau\}.
\nonumber
\end{IEEEeqnarray}
We choose $\tau=-\gamma\log|\mathcal{X}|.$ Then, $\mathcal{A}_\tau = \mathcal{A}_{R,\gamma},$ with the notation of Proposition \ref{prop:half_of_the_one_shot_proof}.
Hence, from \eqref{eq:expected_TV_Q_X_on_shot} and Markov's inequality, we have
\begin{IEEEeqnarray}{c}
Q_{\mathcal{C}}
\Big(
\|Q_{X^{(1)}
|\mathcal{C}} -
p_{X}\|_{TV}
\geq \dfrac{3\Delta}{\varepsilon}[p(\mathcal{A}_{R,\gamma}) + 2^{-\gamma\log(|\mathcal{X}|)/2}]
\Big)
\leq
\dfrac{\varepsilon}{3\Delta}
.
\IEEEeqnarraynumspace\label{eq:TV_Q_X_concentrates_Markov_ineq_one_shot}
\end{IEEEeqnarray}
By construction, we have $Q_{\mathbf{Y}^{(B)},\mathbf{X}^{(B)}} \equiv p_{Y,X}^{\otimes B}.$
Therefore, from \eqref{eq:introducing_R_Delta_B_in_proof_main_prop_one_shot}, and
the additivity of $d,$
we have
$
\mathbb{E}_Q[d(\mathbf{X}^{(B)},\mathbf{Y}^{(B)})] \leq \Delta.
$
Therefore, from
Markov's inequality,
and since
$\varepsilon\in(0,\Delta/2),$
we have
\begin{IEEEeqnarray}{c}
Q_{\mathcal{C}}
\Big(
\mathbb{E}_{Q}[d(\mathbf{X}^{(B)},\mathbf{Y}^{(B)})|\mathcal{C}] 
\geq
\Delta
+ \varepsilon
\Big)
\leq \dfrac{\Delta}{\Delta+\varepsilon}
= 1-\dfrac{\varepsilon}{\Delta}\cdot\dfrac{1}{1+\varepsilon/\Delta}
< 1-\dfrac{2\varepsilon}{3\Delta},
\label{eq:small_proba_of_bad_distortion_one_shot}\IEEEeqnarraynumspace
\end{IEEEeqnarray}
From a union bound and \eqref{eq:average_score_concentrates_Markov_ineq_one_shot}, \eqref{eq:TV_Q_X_concentrates_Markov_ineq_one_shot}, and \eqref{eq:small_proba_of_bad_distortion_one_shot}
there exists a realization $c_*$ of $\mathcal{C}$ such that none of the
corresponding events
hold.
Since, by construction,
$Q_{\mathbf{M}^{(B)},\mathbf{Y}^{(B)},\mathbf{X}^{(B)}|\mathcal{C}{=}c*} \equiv Q_{M^{(1)},Y^{(1)},X^{(1)}|\mathcal{C}{=}c*}^{\otimes B},$
and 
$\delta_{
+}( \mathbf{Y}^{(B)}) \ge \delta( \mathbf{Y}^{(B)})$,
this concludes the proof of Proposition \ref{prop:half_of_the_one_shot_proof}.
\end{IEEEproof}

\subsection{Proof of Theorem \ref{thm:one_shot_achievable_points}}\label{app:subsec_achievability_proof_using_the_existence_of_Q_one_shot}

Fix
some $R>
0
,$
$\Delta>0,$
$\varepsilon \in (0,\Delta/2),$
$\gamma>0,$
and
some
a conditional transition kernel
$p_{Y|X}$
satisfying
$
p_Y \equiv p_X
\text{ and }
\mathbb{E}_p[d(X,Y)] \leq \Delta.
$
Then, we can apply Proposition \ref{prop:half_of_the_one_shot_proof}. We use the notation
therein.

\vspace{5pt}
\subsubsection{Compression scheme achieving close-to-uniform sampling}
We define
$P_{X,M,Y}
=
p_{\scalebox{0.7}{$X$}}
\cdot Q_{\scalebox{0.7}{$M,Y|
X
$}},$
which
differs from $Q$ in having the correct marginal for $X.$
Therefore,
from \eqref{eq:in_main_prop_def_Q_one_shot},
$P$ satisfies the Markov chain
$X{-}M{-}Y.$
Hence,
it
defines an $(R,0)$ code.
Moreover, from \eqref{eq:in_main_prop_def_Q_one_shot}, its decoder $Q_{Y|M}$ is deterministic.
From
Lemma \ref{lemma:TV_same_channel} (Appendix \ref{app:TV_small_batch}), comparing $P$ with $Q$ reduces to comparing marginals,
i.e., to
\eqref{eq:in_main_prop_TV_Q_X_one_shot}
:
\begin{IEEEeqnarray}{rCl}
\IEEEeqnarraymulticol{3}{l}{
\scalebox{0.95}{$
\big\|P_{\scalebox{0.6}{$M, X, Y$}} {-} Q_{\scalebox{0.6}{$M, X, Y
$}}\big\|_{TV}
{=}
\big\|P_{X} {-} Q_{X
}\big\|_{TV}
$}
}
=
\scalebox{0.95}{$
\big\|p_X - Q_{ X
}\big\|_{TV}
\leq
\dfrac{3\Delta}{\varepsilon}[p(\mathcal{A}_{R,\gamma})+2^{-\gamma\log(|\mathcal{X}|)/2}].
$}
\label{eq:TV_P_Q_no_batch_one_shot}
\end{IEEEeqnarray}
Since $d$ is additive, we have
\begin{IEEEeqnarray}{rCl}
\mathbb{E}_{(P)^{\otimes B}}[d(\mathbf{X}^{(B)},\mathbf{Y}^{(B)})] 
=
\mathbb{E}_{P}[d(X,Y)] \ \ \text{and} \ \
\mathbb{E}_{(Q)^{\otimes B}}[d(\mathbf{X}^{(B)},\mathbf{Y}^{(B)})] 
=
\mathbb{E}_{Q}[d(X,Y)].
\nonumber
\end{IEEEeqnarray}
Since $d$ is bounded,
Lemma \ref{lemma:continuity_TV} (Appendix \ref{app:TV_small_batch})
applies.
Then, from
\eqref{eq:TV_P_Q_no_batch_one_shot},
and
Lemma \ref{lemma:TV_joint_to_TV_marginal} with $W=
(X,Y),$
\begin{IEEEeqnarray}{rCl}
\mathbb{E}_{P^{\otimes B}}[d(\mathbf{X}^{(B)},\mathbf{Y}^{(B)})]
&\leq&
\mathbb{E}_{Q^{\otimes B}}[d(\mathbf{X}^{(B)},\mathbf{Y}^{(B)})]
+ \dfrac{6
\Delta}{\varepsilon}\max(d)[p(\mathcal{A}_{R,\gamma})+2^{-\gamma\log(|\mathcal{X}|)/2}]
\nonumber\\*
&\leq&
\Delta + \varepsilon
+ \dfrac{6
\Delta}{\varepsilon}\max(d)[p(\mathcal{A}_{R,\gamma})+2^{-\gamma\log(|\mathcal{X}|)/2}],
\nonumber
\end{IEEEeqnarray}
where the last inequality follows from
\eqref{eq:in_main_prop_distortion_one_shot}.
Moving to the realism performance,
we have the following property of the TVD; see Appendix \ref{app:TV_small_batch}:
\begin{claim}\label{claim:TV_on_small_batch_one_shot}
For any $B\in\mathbb{N}$
and any distributions $P$ and $Q$ on
$\mathcal{X},$
$\|P^{\otimes B}-Q^{\otimes B}\|_{TV} \leq B\|P-Q\|_{TV}.$
\end{claim}

From
Lemma \ref{lemma:continuity_TV}
and the combination of
Claim \ref{claim:TV_on_small_batch_one_shot},
\eqref{eq:TV_P_Q_no_batch_one_shot},
and
Lemma \ref{lemma:TV_joint_to_TV_marginal} with $W=\mathbf{Y}^{(B)},$
we have,
\begin{IEEEeqnarray}{c}
\mathbb{E}_{P^{\otimes B}}[\delta(\mathbf{Y}^{(B)})]
\leq
\mathbb{E}_{Q^{\otimes B}}[\delta(\mathbf{Y}^{(B)})]
+ \dfrac{6B\Delta}{\varepsilon}\max(\delta)[p(\mathcal{A}_{R,\gamma})+2^{-\gamma\log(|\mathcal{X}|)/2}]
\nonumber\\*
\leq
\dfrac{3\Delta}{\varepsilon}
\cdot \left({
1+}
\dfrac{B^2}{\lfloor 2^{R} \rfloor}
\max_x B\log\dfrac{
2}{p_X(x)} \right)
+ \dfrac{6B\Delta}{\varepsilon}[p(\mathcal{A}_{R,\gamma})+2^{-\gamma\log(|\mathcal{X}|)/2}]\cdot \max_x B\log\dfrac{2}{p_X(x)},
\nonumber
\end{IEEEeqnarray}
where the last inequality follows from \eqref{eq:in_main_prop_realism_one_shot} and Claim \ref{claim:upper_bound_of_average_score_uniform_over_all_delta}.
This
concludes the proof.



\section{Achievability of Theorem \ref{thm:small_batch_size_asymptotics}}
\label{app:achievability_small_batch_size}

Fix
$\Delta\in\mathbb{R}_+$ such that $R^{(1)}(\Delta)<H_p(X),$ and
a sequence $\{B_n\}_{n \geq 1}$ of positive integers such that
\begin{IEEEeqnarray}{c}
    \log(B_n)/n \underset{n \to \infty}{\longrightarrow} 0.\label{eq:sub_exponential_growth_at_beginning_of_proof_thm_small_batch_size}
\end{IEEEeqnarray}
We prove that
$R_*(\Delta)\leq R^{(1)}(\Delta).$
Fix $R\in(R^{(1)}(\Delta),H_p(X))$ and $\varepsilon \in \big(0,\min\big(\Delta/2, \ R-R^{(1)}(\Delta) \big)\big),$ and $\gamma\in(0,\varepsilon/\log(|\mathcal{X}|)).$ Then, from \eqref{eq:def_RDP_function}, there exists $p_{Y|X}$ such that
\begin{IEEEeqnarray}{c}
p_Y \equiv p_X, \ \mathbb{E}_p[d(X,Y)] \leq \Delta, \ R \geq I_p(X;Y)+\varepsilon.
\label{eq:introducing_R_Delta_B_in_proof_thm_small_batch_size}
\end{IEEEeqnarray}
We use the powerful result of Theorem \ref{thm:universal_semi-measure_and_critic} regarding the existence of a so-called \textit{universal critic}
$\delta_0.$
From Definition \ref{def:sum_critic_at_beginning_of_paper}, for every $n\in\mathbb{N},$ the restriction of $\delta_0$ to $\mathcal{X}^{nB_n}$ is a $p_X^{\otimes nB_n}$-critic.
Then,
from the additivity of $d,$ we can apply Theorem \ref{thm:one_shot_achievable_points} for set $\mathcal{X}^n,$ distribution $p_X^{\otimes n},$
transition kernel $\prod_{t{=}1}^n p_{Y|X},$
batch size $B_n,$ critic $\delta_0,$ rate $nR
,$ and constants $\Delta,\varepsilon,\gamma.$ This gives that, for every $n$ large enough, there is a $(n,R,0)$ code having a deterministic decoder, inducing a distribution $P^{(n)}$ such that
\begin{IEEEeqnarray}{rCl}
\mathbb{E}_{(P^{(n)})^{\otimes 
B_n
}}
\big[
d(
\mathbf{X}^{(n,
B_n
)}
,
\mathbf{Y}^{(n,
B_n
)}
)
\big]
&\leq&
\Delta + \varepsilon + \dfrac{6\Delta}{\varepsilon}\max(d)[p_{X,Y}^{\otimes n}(\mathcal{A}^{(n)}_{R,\gamma})+2^{-\gamma n \log(|\mathcal{X}|)/2}],
\IEEEeqnarraynumspace
\label{eq:first_big_ineq}
\\
\scalebox{0.97}{$
\mathbb{E}_{(P^{(n)})^{\otimes 
B_n
}}
\big[
\delta_0(
\mathbf{Y}^{(n,
B_n
)}
)
\big]
$}
&\leq&
\scalebox{0.97}{$
\dfrac{3\Delta}{\varepsilon}
\Bigg[ {
1+}
\dfrac{B_n^2}{\lfloor 2^{nR} \rfloor}
\max_x \left( nB_n\log\dfrac{
2}{p_X(x)} \right)
$}
\nonumber
\\*
&&
\scalebox{0.95}{$
+2B_n[p_{X,Y}^{\otimes n}(\mathcal{A}^{(n)}_{R,\gamma})+2^{-\gamma n \log(|\mathcal{X}|)/2}]\cdot \max_x \left( nB_n\log\dfrac{2}{p_X(x)} \right)
\Bigg]
$},
\IEEEeqnarraynumspace
\label{eq:second_big_ineq}
\end{IEEEeqnarray}
\begin{IEEEeqnarray}{c}
\text{where} \quad
\mathcal{A}_{R,\gamma}^{(n)}:= \Big\{(x_{1:n},y_{1:n})\in(\mathcal{X}^n)^2 \ | \  \sum_{t{=}1}^n\log\Big(\dfrac{p_{X,Y}(x_t,y_t)}{p_X(x_t)p_Y(y_t)}\Big)-\log(\lfloor2^{nR}\rfloor) > -\gamma n \log(|\mathcal{X}|) \Big\},
\nonumber
\end{IEEEeqnarray}
with the convention $0/0:=1.$
From \eqref{eq:introducing_R_Delta_B_in_proof_thm_small_batch_size}, $\log(\lfloor2^{nR}\rfloor)/n -\gamma \log(|\mathcal{X}|)>I_p(X;Y)$ for large enough $n.$ Then, since $\mathcal{X}$ is finite, we have, from Hoeffding's inequality,
$
p_{X,Y}^{\otimes n}(\mathcal{A}^{(n)}_{R,\gamma}) = O(e^{-\kappa n}),
$
for some $\kappa>0.$
Hence, from \eqref{eq:sub_exponential_growth_at_beginning_of_proof_thm_small_batch_size},
\eqref{eq:first_big_ineq},
\eqref{eq:second_big_ineq},
and
Theorem \ref{thm:universal_semi-measure_and_critic},
$
\limsup_{n \to \infty}
\mathbb{E}_{(P^{(n)})^{\otimes 
B_n
}}
\big[
d(
\mathbf{X}^{(n,
B_n
)}
,
\mathbf{Y}^{(n,
B_n
)}
)
\big]
\leq \Delta
+\varepsilon,
$
and for any lower semi-computable $p_X^{\otimes *}$-critic $\delta
,$
we have
$
\sup_{n\in\mathbb{N}}
\mathbb{E}_{
(P^{(n)})^{\otimes 
B_n
}
}
\big[
\delta
(\mathbf{Y}^{(n,
B_n
)})
\big]
<\infty
.
$
Since
$P^{(n)}$ has a deterministic decoder,
which we denote
by $m \mapsto y_{1:n}(m),$
it only remains to derandomize
its
encoder.
The claim below is a slight modification of \cite[Prop.~4]{HamdiEtAl2024RDPPrivateRandomness}.
See
the next section
for a proof.
\begin{claim}\label{claim:derandomizing_encoder_gives_exponential_decay_of_additional_TV}
There exists a sequence of deterministic maps
$
f^{(n)}: \mathcal{X}^n
\to [2^{nR}],
$
such that
\begin{IEEEeqnarray}{c}
\scalebox{0.9}{$
\big\|\hat{\Tilde{P}}^{(n)}_{\mathcal{X}^2}[X_{1:n}, y_{1:n}(M)] - \hat{P}^{(n)}_{\mathcal{X}^2}[X_{1:n}, y_{1:n}(M)]\big\|_{TV}
\underset{n \to \infty}{\longrightarrow} 0
,$}
\nonumber
\\*
\scalebox{0.9}{$\liminf_{n\to\infty}\dfrac{-1}{n}\log\big\|\Tilde{P}^{(n)}_{M} - P^{(n)}_{M}\big\|_{TV}
>0
, \quad \text{where}
$}\label{eq:in_claim_encoder_private_randomness_exponential_decay_TV_sur_M} \IEEEeqnarraynumspace
\\
\Tilde{P}^{(n)}_{X_{1:n},M}:=p_{X}^{\otimes n}
\cdot \mathbf{1}_{M=f^{(n)}(X_{1:n})}.\nonumber 
\end{IEEEeqnarray}
\end{claim}
Then, from \eqref{eq:sub_exponential_growth_at_beginning_of_proof_thm_small_batch_size} and Claim \ref{claim:TV_on_small_batch_one_shot}, we have
$
\scalebox{0.9}{$
\liminf_{n\to\infty}
\tfrac{-1}{n}
\log
\|
(\Tilde{P}^{(n)})^{\otimes B_n}_{M}
-
(P^{(n)})^{\otimes B_n}_{M}
\|_{TV}
>0.
$}
$
Thus, from Lemma \ref{lemma:continuity_TV} and \eqref{eq:bound_in_def_sum_critic_if_supported_on_fixed_length_set}, we have
$
|\mathbb{E}_{
(\Tilde{P}^{(n)})^{\otimes 
B_n
}
}
[
\delta
(\mathbf{Y}^{(n,
B_n
)})
]
-
\mathbb{E}_{
(P^{(n)})^{\otimes 
B_n
}
}
[
\delta
(\mathbf{Y}^{(n,
B_n
)})
\big
|
\
\substack{\raisebox{-4pt}{$\longrightarrow$} \\ \scalebox{0.7}{$n$$\to$$\infty$}}
\
0.
$
Moreover, since $d$ is bounded, then from Lemma \ref{lemma:continuity_TV}, we obtain
\begin{IEEEeqnarray}{c}
\Big|
\mathbb{E}_{(P^{(n)})^{\otimes 
B_n
}}
\big[
d(
\mathbf{X}^{(n,
B_n
)}
,
\mathbf{Y}^{(n,
B_n
)}
)
\big]
-
\mathbb{E}_{(P^{(n)})^{\otimes 
B_n
}}
\big[
d(
\mathbf{X}^{(n,
B_n
)}
,
\mathbf{Y}^{(n,
B_n
)}
)
\big]
\Big|
\underset{n \to \infty}{\longrightarrow} 0.
\nonumber
\end{IEEEeqnarray}
Since this analysis is valid for any $\varepsilon \in \big(0,\min\big(\Delta/2, \ R-R^{(1)}(\Delta)\big)\big),$ then the tuple $(R,\{B_n\}_{n\geq 1},\Delta)$ is asymptotically achievable with algorithmic realism with fully deterministic codes. This being true for any $R\in(R^{(1)}(\Delta),H_p(X)),$ we have
$
R_*(\Delta)\leq R^{(1)}(\Delta),$
as desired.

\subsection{Encoder derandomization}\label{app:subsec_proof_derandomization}

We show that Claim \ref{claim:derandomizing_encoder_gives_exponential_decay_of_additional_TV}
follows from \cite[Proposition~4]{HamdiEtAl2024RDPPrivateRandomness} and its proof.
We can apply that result directly, since $R<H_p(X)$ and $\mathcal{X}$ is finite. This would give all properties in Claim \ref{claim:derandomizing_encoder_gives_exponential_decay_of_additional_TV}, except for the exponential decay in \eqref{eq:in_claim_encoder_private_randomness_exponential_decay_TV_sur_M}.
To obtain the latter, it is sufficient to adapt the proof of \cite[Proposition~4]{HamdiEtAl2024RDPPrivateRandomness}, by replacing the use of the law of large numbers with the use of Hoeffding's inequality, and using \cite[Theorem~VII.1]{2013PaulCuffDistributedChannelSynthesis} with $\tau=-n\gamma,$ for small enough $\gamma.$

\section{Converse of Theorem \ref{thm:small_batch_size_asymptotics}}\label{app:converse}

From standard information-theoretic arguments, we have the result below. See Appendix \ref{app:standard_converse_arguments} for a proof.
\begin{lemma}\label{lemma:standard_converse_arguments}
Consider a triplet $(R,R_c,\Delta)$ and a sequence of codes, the $n$-th being $(n,R,R_c),$ inducing a sequence $\{P^{(n)}_{X_{1:n},J,M,Y_{1:n}}\}_{n \geq 1}$ of distributions such that
\begin{IEEEeqnarray}{c}
\limsup_{n \to \infty}
\mathbb{E}_{(P^{(n)})^{\otimes b_n}}
\big[
d(
\mathbf{X}^{(n,b_n)}
,
\mathbf{Y}^{(n,b_n)}
)
\big]
\leq \Delta,
\label{eq:in_lemma_in_converse_distortion_constraint}
\IEEEeqnarraynumspace
\end{IEEEeqnarray}
for some sequence $\{b_n\}_{n \geq 1}$ of positive integers. For every $n \geq 1,$ let $T^{(n)}$ denote a uniform variable on $[nb_n]$ independent from
all other random variables. Then, there exists a conditional distribution $p_{Y|X}$ and an increasing sequence $\{n_i\}_{i \geq 1}$ of positive integers such that
\begin{IEEEeqnarray}{c}
(P^{(n_i)})^{\otimes b_{n_i}}_{X_{T^{(n_i)}},Y_{T^{(n_i)}}} \underset{i\to\infty}{\longrightarrow} p_{X,Y}
,\quad
\Delta \geq \mathbb{E}_p[d(X,Y)]
,\quad
R \geq I_p(X;Y).
\nonumber
\end{IEEEeqnarray}
\end{lemma}

\subsection{Converse proof of Theorem \ref{thm:small_batch_size_asymptotics}}

Fix
$\Delta\in\mathbb{R}_+$ such that $R^{(1)}(\Delta)<H_p(X),$ and
a sequence $\{B_n\}_{n \geq 1}$ of positive integers such that
$
\log(B_n)/n
\ \
\substack{\raisebox{-4pt}{$\longrightarrow$} \\ \scalebox{0.7}{$n$$\to$$\infty$}}
\ \
0.
$
We know that $R_*(\Delta) \geq R(\Delta),$ and prove that $R(\Delta) \geq R^{(1)}(\Delta).$
Consider
$
R
\in \mathbb{R}_+
,$ and some $R_c \in \mathbb{R}_{\geq 0}$ such that
$(R, R_c, \{B_n\}_{n \geq 1}, \Delta)$ is asymptotically achievable with
algorithmic
realism. Fix $\varepsilon>0.$
Then, there exists
a
sequence of codes,
the $n$-th being
$(n,R,
R_c),$
inducing a sequence 
$\{P^{(n)}_{X_{1:n},J,M,Y_{1:n}}\}_{n\in\mathbb{N}}$
of distributions
such that
$
\limsup_{n \to \infty}
\mathbb{E}_{(P^{(n)})^{\otimes 
B_n
}}
\big[
d(
\mathbf{X}^{(n,
B_n
)}
,
\mathbf{Y}^{(n,
B_n
)}
)
\big]
\leq \Delta+\varepsilon,
$
and for any lower semi-computable $p_X^{\otimes *}$-critic $\delta,$
\begin{IEEEeqnarray}{c}
\sup_{n\in\mathbb{N}}
\mathbb{E}_{
(P^{(n)})^{\otimes 
B_n
}
}
\big[
\delta
(\mathbf{Y}^{(n,
B_n
)})
\big]
<\infty
.
\label{eq:in_converse_distributional_realsim_property}
\end{IEEEeqnarray}
Then, Lemma \ref{lemma:standard_converse_arguments} applies, with $b_n=B_n,$ for all $n,$ with $R+\varepsilon$ instead of $R,$ and $\Delta+\varepsilon$ instead of $\Delta.$ Then, there exists
$p_{Y|X}$ and an increasing sequence $\{n_i\}_{i \geq 1}$ of positive integers such that
\begin{IEEEeqnarray}{c}
    (P^{(n_i)})^{\otimes b_{n_i}}_{X_{T^{(n_i)}},Y_{T^{(n_i)}}} \underset{i\to\infty}{\longrightarrow} p_{X,Y}\label{eq:in_converse_small_batch_size_converging_subsequence}
    \\
    \Delta+\varepsilon \geq \mathbb{E}_p[d(X,Y)]
    \label{eq:in_converse_small_batch_size_single_letter_distortion_bound}
    \\
    R+\varepsilon \geq I_p(X;Y),\label{eq:in_converse_small_batch_size_single_letter_rate_bound}
\end{IEEEeqnarray}
where for any $n\in\mathbb{N},$ variable $T^{(n)}$ is uniformly distributed on $[nB_n],$ and independent from 
all other random variables.
We
prove that $p_Y \equiv p_X.$
Fix $e_0 \in
\mathcal{X}
.$
Consider the computable $p_X^{\otimes *}$-critic $\delta$ from Proposition~\ref{prop:frequency_critic_any_alphabet}, with $q$ therein taken to be $p_X.$
Then, from \eqref{eq:in_converse_distributional_realsim_property},
\begin{IEEEeqnarray}{c}
\sup_{n\in\mathbb{N}}
\mathbb{E}_{
(P^{(n)})^{\otimes 
B_n
}
}
\big[
\delta
(\mathbf{Y}^{(n,
B_n
)})
-2\log(
\delta
(\mathbf{Y}^{(n,
B_n
)})
+
3
)
\big]
<\infty
.
\nonumber
\end{IEEEeqnarray}
\begin{IEEEeqnarray}{c}
\text{Thus,}\quad
\sup_{n\in\mathbb{N}}
\mathbb{E}_{
(P^{(n)})^{\otimes 
B_n
}
}
\big[
\delta
(\mathbf{Y}^{(n,
B_n
)})
\big]
<\infty
,
\ 
\text{ and } \
\mathbb{E}_{
(P^{(n)})^{\otimes 
B_n
}
}
\big[
\delta
(\mathbf{Y}^{(n,
B_n
)})
-\dfrac{1}{2}\log(nB_n)
\big]
\underset{n\to\infty}{\longrightarrow} -\infty.
\nonumber
\end{IEEEeqnarray}
Therefore,
the frequency of
$e_0$
in a batch of reconstructions
converges in $L_1$ norm to $p_X(
e_0
).$
Hence, the expected frequencies converge to $p_X(
e_0
).$ This rewrites as
$
(P^{(n)})^{\otimes B_{n}}_{Y_{T^{(n)}}}(
e_0
) \to p_X(
e_0
).
$
This is true for any $e_0$ in $
\mathcal{X}
.$
Thus, from \eqref{eq:in_converse_small_batch_size_converging_subsequence}, $p_Y \equiv p_X.$ Hence, from \eqref{eq:in_converse_small_batch_size_single_letter_distortion_bound} and \eqref{eq:in_converse_small_batch_size_single_letter_rate_bound}, we have
$
R+\varepsilon \geq R^{(1)}(\Delta+\varepsilon).
$
This being true for any $\varepsilon>0,$ and since $R^{(1)}(\cdot)$ is convex ---
thus continuous
--- on $(0,\infty),$ we have
$
R \geq R^{(1)}(\Delta).
$
This being true for any $R\in\mathbb{R}_+$ such that there exists $R_c \in \mathbb{R}_{\geq 0}$ for which
$(R, R_c, \{B_n\}_{n \geq 1}, \Delta)$ is asymptotically achievable with
algorithmic
realism, we have
$
R(\Delta)\geq R^{(1)}(\Delta),
$
as desired.
\section{Proof of Theorem \ref{thm:very_large_batch_size}}
\label{app:proof_thm_large_batch_size_asymptotics}
Consider
an
increasing
sequence $\{B_n\}_{n \geq 1}
$ of positive integers such that
$
B_n
/
|\mathcal{X}|^n
\
\substack{\raisebox{-4pt}{$\longrightarrow$} \\ \scalebox{0.7}{$n$$\to$$\infty$}}
\
\infty,
$
some $R_c \in \mathbb{R}_{\geq 0},$ and some $(R,\Delta)\in(\mathbb{R}_+)^2$ such that tuple $(R, R_c,
\Delta)$ is asymptotically achievable with near-perfect realism.
From Theorem 1 in \cite{2022AaronWagnerRDPTradeoffTheRoleOfCommonRandomness}, $(R,R_c,\Delta)$
is also
achievable with
perfect realism
(Section~\ref{sec:short_background_on_distribution_matching}).
Fix $\varepsilon>0,$
and a corresponding sequence
$\{(F^{(n)},G^{(n)})\}_n$
of codes,
the $n$-th being
$(n,R+\varepsilon,R_c)$ 
\footnote{The fact that there is no need for an $\varepsilon$ backoff in the common randomness rate follows from the proof of \cite[Theorem~1]{2022AaronWagnerRDPTradeoffTheRoleOfCommonRandomness}.}. Denote by $P^{(n)}$ the distribution induced by the $n$-th code.
Then,
there exists an integer $N_\varepsilon$ such that
\begin{IEEEeqnarray}{c}
\limsup_{n \to \infty}
\mathbb{E}_{
P^{(n)}
}
\big[
d(
X_{1:n}
,
Y_{1:n}
)
\big]
\leq \Delta
+\varepsilon,
\quad \text{and} \quad
\scalebox{1.0}{$\forall n\geq N_\varepsilon,
\ 
P^{(n)}_{Y_{1:n}}
\equiv p_{X}^{\otimes n
}.
$}
\label{eq:in_converse_large_batch_size_perfect_realsim_distortion_constraint}
\end{IEEEeqnarray}
\begin{IEEEeqnarray}{c}
\text{Thus,}\quad
\scalebox{1.0}{$\forall n\geq N_\varepsilon,
\ (P^{(n)}_{Y_{1:n}})^{\otimes
B_n
} \equiv p_{X}^{\otimes n
B_n
}.
$}\label{eq:batch_perfect_realism}
\end{IEEEeqnarray}
From \eqref{eq:in_converse_large_batch_size_perfect_realsim_distortion_constraint}, \eqref{eq:batch_perfect_realism}, Claim \ref{claim:upper_bound_of_average_score_uniform_over_all_delta}, and the additivity of the distortion measure $d,$ we have
\begin{IEEEeqnarray}{c}
\limsup_{n \to \infty}
\mathbb{E}_{(P^{(n)})^{\otimes 
B_n
}}
\big[
d(
\mathbf{X}^{(n,
B_n
)}
,
\mathbf{Y}^{(n,
B_n
)}
)
\big]
\leq \Delta
+\varepsilon,
\nonumber
\end{IEEEeqnarray}
and for any lower semi-computable $p_X^{\otimes *}$-critic $\delta
,$
we have
$
\sup_{n\in\mathbb{N}}
\mathbb{E}_{
(P^{(n)})^{\otimes 
B_n
}
}
[
\delta
(\mathbf{Y}^{(n,
B_n
)})
]
<\infty
.
$
Since this analysis is valid for every $\varepsilon>0,$
then
$(R,R_c,
\{B_n\}_{n\geq 1},
\Delta)
$
is asymptotically achievable with algorithmic realism.
Moving to the converse,
fix
a computable
increasing sequence
$\{B_n\}_{n \geq 1}
$
of positive integers
with
\begin{IEEEeqnarray}{c}
    \dfrac{B_n}{|\mathcal{X}|^n
    } \to \infty,\label{eq:assumption_on_B_n_in_proof_very_large_batch_size}
\end{IEEEeqnarray}some $R_c \in \mathbb{R}_{\geq 0},$ and some $(R,\Delta)\in(\mathbb{R}_+)^2$ such that tuple $(R, R_c,
\Delta)$ is asymptotically achievable with algorithmic realism.
Fix $\varepsilon>0.$
Then,
there exists
a sequence of codes, the $n$-th being $(n, R
+\varepsilon, 
R_c),$
such that the sequence $\{P^{(n)}\}_n$ of distributions induced by the codes satisfies
\begin{IEEEeqnarray}{c}
\limsup_{n \to \infty}
\mathbb{E}_{(P^{(n)})^{\otimes 
B_n
}}
\big[
d(
\mathbf{X}^{(n,
B_n
)}
,
\mathbf{Y}^{(n,
B_n
)}
)
\big]
\leq \Delta
+\varepsilon,
\label{eq:batch_distributional_realsim_distortion_in_proof_thm_very_large_batch}
\IEEEeqnarraynumspace
\end{IEEEeqnarray}
and for any lower semi-computable $p_X^{\otimes *}$-critic $\delta,$
\begin{IEEEeqnarray}{c}
\sup_{n\in\mathbb{N}}
\mathbb{E}_{
(P^{(n)})^{\otimes 
B_n
}
}
\big[
\delta
(\mathbf{Y}^{(n,
B_n
)})
\big]
<\infty
.
\label{eq:distributional_realsim_perception_in_proof_thm_very_large_batch}
\IEEEeqnarraynumspace
\end{IEEEeqnarray}
\begin{lemma}\label{lemma:sample_complexity_of_approximating_a_discrete_distribution}\cite{SampleComplexityEstimatingDiscreteDistributions}
There exists
a positive integer
$\lambda$
such that, for any $k \in \mathbb{N},$ any distribution $q$ on some finite set $\mathcal{W}$ of size $k,$ any $\varepsilon,\eta > 0,$ and any integer $b$ satisfying
$
b \geq \lambda \cdot
[
k + \log(1/\eta)
]
/
\varepsilon^2
,
$
we have
\begin{IEEEeqnarray}{c}
q^{\otimes b}\Big(
\big\|
\mathbb{P}^{\text{emp}}_{\mathcal{W}}
[W^b]
- q
\big\|_{TV}
\geq \varepsilon
\Big)
\leq \eta
.
\nonumber
\end{IEEEeqnarray}
\end{lemma}
For every $n\in\mathbb{N},$ define
\begin{IEEEeqnarray}{c}
C_n := \Bigg\lceil \bigg( \dfrac{B_n}{|\mathcal{X}|^n} \bigg)^{\tfrac{1}{3}} \Bigg\rceil.\label{eq:def_C_n}
\end{IEEEeqnarray}
Since $\mathcal{X}$ is finite, $\{C_n\}_{n \geq 1}$ is a computable sequence of positive integers. Moreover,
from \eqref{eq:assumption_on_B_n_in_proof_very_large_batch_size},
we have
\begin{IEEEeqnarray}{c}
    C_n \underset{n \to \infty}{\longrightarrow} \infty
    .\label{eq:assumptions_on_C_n_and_B_n_in_proof_very_large_batch_size}
\end{IEEEeqnarray}
Choosing, for every $n\in\mathbb{N},$ $\eta = 1/3$ and $\varepsilon = 1/C_n,$ then from
Lemma \ref{lemma:sample_complexity_of_approximating_a_discrete_distribution} and
\eqref{eq:assumptions_on_C_n_and_B_n_in_proof_very_large_batch_size} we have
\begin{IEEEeqnarray}{c}
(P^{(n)})^{\otimes 
B_n
}
\Big(
\big\|
\mathbb{P}^{\text{emp}}_{\mathcal{X}^n}
[\mathbf{Y}^{(n,B_n)}]
- P^{(n)}_{Y_{1:n}}
\big\|_{TV}
\geq \dfrac{1}{C_n}
\Big)
\leq \dfrac{1}{3}
,
\IEEEeqnarraynumspace
\label{eq:distribution_estimation_of_P_Y_n}
\end{IEEEeqnarray}
for large enough $n.$
Consider the computable sequence of positive integers defined by
\begin{IEEEeqnarray}{c}
\forall n \in \mathbb{N}, \ A_n:= \Bigg\lceil \bigg( \dfrac{B_n}{|\mathcal{X}|^n} \bigg)^{\tfrac{4}{9}} \Bigg\rceil.
\label{eq:def_A_n}
\end{IEEEeqnarray}
Since $\{B_n\}_{n\geq 1}$ is increasing, then for any $t \in \mathbb{N},$ there exists a unique integer $n\in\mathbb{N}_{\geq 0}$ such that
$t\in[nB_n, (n+1)B_{n+1}),$
with
$B_0:=0.$
We define $\delta:
\cup_{t\in\mathbb{N}} \mathcal{X}^t
\to \mathbb{N}_{\geq 0}$ as follows.
For any integer $t \in [1,B_1),$ and
$x_{1:t} \in
\mathcal{X}
^t,$ let $\delta(x):=0.$
For any $n \in \mathbb{N},$
$t\in[nB_n, (n+1)B_{n+1}),$ and
$x_{1:t} \in
\mathcal{X}
^t,$ let
\begin{IEEEeqnarray}{c}
\delta(x_{1:t}):= \Big
\lceil
A_n 
\big\|
\mathbb{P}^{\text{emp}}_{\mathcal{X}^n}
[x_{1:n
B_n
}]
- p_X^{\otimes n}
\big\|_{TV}
\Big
\rceil
.
\nonumber
\end{IEEEeqnarray}
\begin{claim}\label{claim:TV_based_critic}
There exists
$L\in\mathbb{N}$ such that $\delta-2\log(\delta+
3
)-
L
$ is a
lower semi-computable
$p_X^{\otimes *}$-critic.
\end{claim}

We provide a proof in Appendix \ref{app:subsec:proof_claim_TV_based_critic}. Then, we can apply \eqref{eq:distributional_realsim_perception_in_proof_thm_very_large_batch} to critic $\delta-2\log(\delta+
3
)-
L
,$
and get,
\begin{IEEEeqnarray}{c}
\sup_{n\in\mathbb{N}}
\mathbb{E}_{
(P^{(n)})^{\otimes 
B_n
}
}
\big[
\delta
(\mathbf{Y}^{(n,
B_n
)})
-2\log(
\delta
(\mathbf{Y}^{(n,
B_n
)})
+
3
)
-
L
\big]
<\infty
.
\nonumber
\end{IEEEeqnarray}
\begin{IEEEeqnarray}{c}
\text{Thus,}\quad
\sup_{n\in\mathbb{N}}
\mathbb{E}_{
(P^{(n)})^{\otimes 
B_n
}
}
\big[
\delta
(\mathbf{Y}^{(n,
B_n
)})
\big]
<\infty
,
\ 
\text{ and } \
(P^{(n)})^{\otimes 
B_n
}
\big(
\delta
(\mathbf{Y}^{(n,
B_n
)})
\geq C
_n
\big)
\underset{n\to\infty}{\longrightarrow} 0,
\nonumber
\end{IEEEeqnarray}because $\{C_n\}_{n\geq 1}$ tends to infinity.
Combining this with \eqref{eq:distribution_estimation_of_P_Y_n} through a union bound, we obtain, from the triangle inequality for the TVD:
for large enough $n,$
\begin{IEEEeqnarray}{c}
(P^{(n)})^{\otimes 
B_n
}
\bigg(
\big\|
P^{(n)}_{Y_{1:n}}
- p_X^{\otimes n}
\big\|_{TV}
\leq
\dfrac{C_n}{A_n}
+
\dfrac{1}{C_n}
\bigg)
>0
.
\nonumber
\end{IEEEeqnarray}
The above event does not depend on the random batch, hence the corresponding inequality is true, for large enough $n.$ Since $\{C_n\}_{n \geq 1}$ tends to infinity and since from
\eqref{eq:assumption_on_B_n_in_proof_very_large_batch_size}, \eqref{eq:def_C_n},
and \eqref{eq:def_A_n},
we have $C_n/A_n \to 0,$ then we obtain
$
\|
P^{(n)}_{Y_{1:n}}
- p_X^{\otimes n}
\|_{TV}
\ \
\substack{\raisebox{-4pt}{$\longrightarrow$} \\ \scalebox{0.7}{$n$$\to$$\infty$}}
\ \
0.
$
Hence,
from \eqref{eq:batch_distributional_realsim_distortion_in_proof_thm_very_large_batch} and the additivity of $d,$ we have that $(R,R_c,\Delta)$ is asymptotically achievable with near-perfect realism.
This concludes the proof.



\section{Standard converse arguments}\label{app:standard_converse_arguments}

Here, we provide a proof of Lemma \ref{lemma:standard_converse_arguments} (Appendix \ref{app:converse}).
As mentioned in Section \ref{sec:background},
we have $\mathcal{X}\subseteq \{0,1\}^s$ for some positive integer $s.$
Then,
the sequence of
distributions
$(P^{(n)})^{\otimes b_n}_{X_T,Y_T}$ can be seen as a bounded sequence in $\mathbb{R}^{2^{2s}},$ thus it admits a converging subsequence:
\begin{IEEEeqnarray}{c}
    (P^{(n_i)})^{\otimes b_n}_{X_T,Y_T} \underset{i\to\infty}{\longrightarrow} p_{X,Y}.\label{eq:converging_subsequence}
\end{IEEEeqnarray}
Since $d$ is bounded, we have
\begin{IEEEeqnarray}{c}
    \mathbb{E}_{(P^{(n_i)})^{\otimes b_n}}[d(X_T,Y_T)] \underset{i\to\infty}{\longrightarrow} \mathbb{E}_{p}[d(X,Y)].\label{eq:distortion_for_converging_subsequence}
\end{IEEEeqnarray}
Since $d$ is additive, we have, for any $n \in \mathbb{N},$
\begin{IEEEeqnarray}{c}
\mathbb{E}_{(P^{(n)})^{\otimes b_n}}
\big[
d(
\mathbf{X}^{(n,b_n)}
,
\mathbf{Y}^{(n,b_n)}
)
\big]
=
\mathbb{E}_{(P^{(n)})^{\otimes b_n}}
\big[
d(
X_T
,
Y_T
)
\big].
\label{eq:same_distortion}
\end{IEEEeqnarray}
From \eqref{eq:in_lemma_in_converse_distortion_constraint}, \eqref{eq:distortion_for_converging_subsequence} and \eqref{eq:same_distortion}, we have
$
\Delta \geq \mathbb{E}_p[d(X,Y)].
$
Secondly, distribution $P^{(n)}$ satisfies
\begin{IEEEeqnarray}{rCl}
    nb_nR
    \geq
    H(\{m^{(k)}\}_{k\in[b_n]}|\{J^{(k)}\}_{k\in[b_n]})
    &\geq&
    I(\{m^{(k)}\}_{k\in[b_n]};\mathbf{X}^{(n,b_n)}|\{J^{(k)}\}_{k\in[b_n]})
    \nonumber\\
    &=&I(\{m^{(k)}\}_{k\in[b_n]},\{J^{(k)}\}_{k\in[b_n]};\mathbf{X}^{(n,b_n)})\nonumber\\
    &\geq&I(\mathbf{Y}^{(n,b_n)};\mathbf{X}^{(n,b_n)})\nonumber\\
    &\geq& \sum_{k=1}^{b_n} \sum_{t=1}^n I(Y^{(k)}_t;X^{(k)}_t)\nonumber\\
    &=& n b_n I(Y_T;X_T|T)\nonumber\\
    &=& n b_n I(T,Y_T;X_T)\nonumber\\
    &\geq& n b_n I(Y_T;X_T).\nonumber
\end{IEEEeqnarray}
Therefore, from \eqref{eq:converging_subsequence}, and by continuity of mutual information on the set of distributions on $(\{0,1\}^s)^2,$ we have $R \geq I_p(X;Y).$

\section{Special
critics}\label{app:frequency_critic}
\subsection{Critic involving the frequency of a specific pattern}

We provide a proof of Proposition~\ref{prop:frequency_critic_any_alphabet}.

\vspace{5pt}
\subsubsection{$\delta$ is a computable integer-valued map}
The function $S$ is a computable integer-valued map. Since $q(e_0)$ and $1/n$ are computable, then the map inside the inner ceiling function is a computable real-valued map. Denote it by $f.$ 
It is straightforward that if a real-valued map $g$ is computable, and the set $\{x \ | \ g(x)\in\mathbb{Z}\}$ is computable, then $\lceil g \rceil \mathbf{1}_{g \neq 0}$ is a computable integer-valued function.
The set $\{x \ | \ f(x)\in\mathbb{Z}\}$ is computable.
Indeed, this set is empty if there does not exist integers $a,b,a',b',c$ such that $q(e_0)=a/b - a'\sqrt{c}/b'.$
Otherwise,
there exists an algorithm having such a tuple $(a,b,a',b',c)$ in memory, which always terminates and determines whether $f(x)$ is rational by computing prime decompositions.
Then,
$\lceil f \rceil \mathbf{1}_{f \neq 0}$ is a computable integer-valued function.
Thus, for the logarithm in base 2 (or any positive integer), $\{x \ | \ \mathbf{1}_{f(x) \neq 0}\log(\lceil f(x) \rceil) \in \mathbb{Z}\}$ is computable.
Moreover, from point (ii) of Lemma \ref{lemma:semicomputability_preserved_by_certain_operations}, $\mathbf{1}_{f \neq 0}\log(\lceil f \rceil)$ is a real-valued computable map.
Then, we can apply the same principle once more with $g=\mathbf{1}_{f \neq 0}\log(\lceil f \rceil).$
Therefore,
$\delta$ is a computable integer-valued map.

\vspace{5pt}
\subsubsection{Remainder of the proof of
Proposition~\ref{prop:frequency_critic_any_alphabet}}
This implies that $\delta,$ seen as a real-valued map, is computable.
Therefore,
from point (ii) of Lemma \ref{lemma:semicomputability_preserved_by_certain_operations},
$\delta-2\log(\delta+
3
)
$ is a real-valued computable map.
For any $(n,C) \in \mathbb{N}^2,$ and any $x_{1:n
} \in
\mathcal{X}
^n
,$ we have:
\begin{IEEEeqnarray}{rCl}
\{\delta(x_{1:n
}) \geq C\}
&=&
\Big\{ \Big
\lceil
\log\Big\lceil
|S(x_{1:n
})-nq(e_0)| \ \big/ \ \sqrt{n}
\Big\rceil \Big
\rceil
\geq C \Big\}
\nonumber\\*
&=&
\Big\{ \log\Big\lceil
|S(x_{1:n
})-nq(e_0)| \ \big/ \ \sqrt{n}
\Big\rceil
> C-1
\Big\}
\nonumber
\\*
&=&
\Big\{ \Big\lceil
|S(x_{1:n
})-nq(e_0)| \ \big/ \ \sqrt{n}
\Big\rceil
> 2^{C-1}
\Big\}
\nonumber\\*
&=&
\Big\{
|S(x_{1:n
})-nq(e_0)| \ \big/ \ \sqrt{n}
>
2^{C-1}
\Big\}.
\nonumber
\end{IEEEeqnarray}
From this and
Chebyshev inequality, we obtain:
\begin{IEEEeqnarray}{rCl}
q^{\otimes n}(\delta(X_{1:n
}) \geq C)
&\leq&
\mathbb{E}_{q^{\otimes n}}\big[\big(S(X_{1:n
})-
n
q(e_0)
\big)^2/
n
\big] /
4^{C{-}1}
\nonumber\\*
&=&
(q(e_0){-}q(e_0)^2)
/
4^{C{-}1}
\label{eq:binomial_variance}
\\*
&\leq&
4^{-C}.
\nonumber
\end{IEEEeqnarray}where \eqref{eq:binomial_variance} comes from the fact that $S(X_{1:n
})$ follows a binomial distribution $B(n,q(e_0))$.
Therefore,
$
\mathbb{E}_{
q^{\otimes n}}
[
\mathbf{1}_
{\delta(X_{1:n
}) = C}
]
\leq 4^{-C}
\leq 2^{-C}.
$
Thus,
\begin{IEEEeqnarray}{c}
\mathbb{E}_{
q^{\otimes n}}
[
\mathbf{1}_
{\delta(X_{1:n
}) = C}
\cdot 2^{\delta(X_{1:n
})
-2\log(\delta(X_{1:n
})
+
3
)
}
]
\leq \dfrac{1}{
(C+
3
)^2}.
\nonumber
\end{IEEEeqnarray}
This also holds for $C=0.$ Summing over $C \in \mathbb{N}_{\geq 0}$ gives, for any $n\in\mathbb{N},$
\begin{IEEEeqnarray}{c}
\sum_{x_{1:n}\in
\mathcal{X}
^n
}
q^{\otimes n}(
x_{1:n
}) \cdot 2^{\delta(x_{1:n
})
-2\log(\delta(x_{1:n
})
+3
)
} \leq 1.
\nonumber
\end{IEEEeqnarray}
Hence,
we have that $\delta-2\log(\delta
+
3
)
$ is a
computable
$q^{\otimes *}$-critic.

\subsection{Critic involving an empirical distribution}\label{app:subsec:proof_claim_TV_based_critic}

We provide a proof of Claim \ref{claim:TV_based_critic}.
We start with a computability analysis. 
\begin{claim}
The map $f:\cup_{t\in\mathbb{N}} \mathcal{X}^t \to \mathbb{R}$ defined by $\forall t \in [1,B_1)\cap\mathbb{N}, \forall x_{1:t} \in \mathcal{X}^t, f(x_{1:t}):=0,$ and
\begin{IEEEeqnarray}{c}\label{eq:intermediate_def_of_TV_test}
\forall n \in \mathbb{N}, \forall t\in [nB_n, (n+1)B_{n+1})\cap\mathbb{N}, \forall x_{1:t} \in \mathcal{X}^t, \ f(x_{1:t}) := \big\|
\mathbb{P}^{\text{emp}}_{\mathcal{X}^n}
[x_{1:n
B_n
}]
- p^{\otimes n}
\big\|_{TV}\IEEEeqnarraynumspace
\end{IEEEeqnarray}
is computable.
\end{claim}
\begin{IEEEproof}
As mentioned in Section \ref{sec:background},
there exists $s \in \mathbb{N}$ such that $\mathcal{X} \subseteq \{0,1\}^s.$
Then, given some $x\in \cup_{t\in\mathbb{N}} \mathcal{X}^t \to \mathbb{R},$ one can compute the unique corresponding $t$ via a Turing machine. Moreover, since $\{B_n\}_{n \geq 1}$ is computable, one can further compute the unique $n$ such that $t\in [nB_n, (n+1)B_{n+1})$ via a Turing machine, as well as the empirical probability appearing in \eqref{eq:intermediate_def_of_TV_test}.
Since $p$ is a computable distribution, then,
from Lemma \ref{lemma:semicomputability_preserved_by_certain_operations}
(regarding
product, sum, and absolute value),
$f$ is computable.
\end{IEEEproof}

We know that $\{A_n\}_{n \geq 1}$ is computable. From Lemma \ref{lemma:semicomputability_preserved_by_certain_operations}, the product of two computable functions is computable, thus lower semi-computable, and the ceiling function preserves semi-computability. Therefore, $\delta$ is lower semi-computable. Since $\delta - 2\log(\delta+3) = \log(2^\delta/(\delta+3)^2),$ then by Lemma \ref{lemma:semicomputability_preserved_by_certain_operations},
for any positive integer $L,$ function $\delta-2\log(\delta+
3
)-
L
$ is lower semi-computable.

It remains to prove that a certain choice of $L$ yields a $p_X^{\otimes *}$-critic.
From 
\eqref{eq:assumption_on_B_n_in_proof_very_large_batch_size}
and \eqref{eq:def_A_n},
we have
\begin{IEEEeqnarray}{c}\label{eq:minimal_sample_complexity_satisfied}
\exists
N_0 \in \mathbb{N},
\forall n \geq N_0, \ B_n \geq \lambda (|\mathcal{X}|^n + 2) A_n^2.
\end{IEEEeqnarray}
For any integers $n \geq N_0,$ $C \geq 2,$
and
$t\in[nB_n, (n+1)B_{n+1}),$ and any $x_{1:t} \in
\mathcal{X}
^t,$
we have:
\begin{IEEEeqnarray}{rCl}
\{\delta(x_{1:t
}) \geq C\}
&=&
\Big\{
\Big
\lceil
A_n 
\big\|
\mathbb{P}^{\text{emp}}_{\mathcal{X}^n}
[x_{1:n
B_n
}]
- p^{\otimes n}
\big\|_{TV}
\Big
\rceil
\geq C \Big\}
=
\Big\{
A_n 
\big\|
\mathbb{P}^{\text{emp}}_{\mathcal{X}^n}
[x_{1:n
B_n
}]
- p^{\otimes n}
\big\|_{TV}
>
C-1 \Big\}.
\nonumber
\end{IEEEeqnarray}
From this,
\eqref{eq:minimal_sample_complexity_satisfied},
and
Lemma \ref{lemma:sample_complexity_of_approximating_a_discrete_distribution},
with distribution
$p_X^{\otimes n},$
and
$b=B_n,$
$
\varepsilon=(C{-}1)/A_n,$
$\eta=2^{-C},$
we have,
$
p_X^{\otimes t}(\delta(X_{1:t
}) \geq C)
\leq
2^{-C},
$
for any
$
t \geq [N_0 B_{N_0},\infty)\cap\mathbb{N},$
and
$
C \in \mathbb{N}_{\geq 2}.
$
Therefore,
\begin{IEEEeqnarray}{c}
\forall t \geq [N_0 B_{N_0},\infty)\cap\mathbb{N}, \forall C \in \mathbb{N}_{\geq 2},
\quad
\mathbb{E}_{p_X^{\otimes t}}[
\mathbf{1}_{
\delta(X_{1:t
}) = C
}
\cdot 2^{\delta(X_{1:t
})
-2\log(\delta(X_{1:t
})
+
3
)
-
1
}
]
\leq \dfrac{1}{
(C+
3
)^2}.\nonumber
\end{IEEEeqnarray}
This also holds for $C\in\{0,1\}.$ Summing over $C \in \mathbb{N}_{\geq 0}$ gives,
\begin{IEEEeqnarray}{c}
\forall t \geq [N_0 B_{N_0},\infty)\cap\mathbb{N},
\quad
\sum_{x_{1:t}\in
\mathcal{X}
^t
}
p_X^{\otimes t}(
x_{1:t
}) \cdot 2^{\delta(x_{1:t
})
-2\log(\delta(x_{1:t
})
+
3
)
-1
} \leq 1.
\nonumber
\end{IEEEeqnarray}
To extend this to all
$t
\in\mathbb{N}
,$
it is sufficient to multiply by $2^{-L}$ for some $L$ large enough.
Thus,
there exists $L \in \mathbb{N}$ such
that $\delta-2\log(\delta+
3
)-L$ is a lower semi-computable $p_X^{\otimes *}$-critic. This concludes.

\subsection{Proof of Proposition \ref{prop:critic_longest_run} }\label{app:subsec:proof_Erdos_Renyi}

Note that $x \mapsto q,$
$x \mapsto l(x),$ and $x \mapsto R(x)$ are computable maps from $\cup_{n\in\mathbb{N}} \mathcal{X}^n$ to $\mathbb{R}.$ Then, from
part (ii) of
Lemma~\ref{lemma:semicomputability_preserved_by_certain_operations},
$x \mapsto \log(q)$ and $x \mapsto \log(l(x)),$
and thus
$x \mapsto \log_{1/q}(l(x))$
are computable.
Define
\begin{IEEEeqnarray}{c}
\forall n\in\mathbb{N}, \
a_n := \log\left(1+\sqrt{\mathrm{Var}(R(X_{1:n}))} + | \mathbb{E}[R(X_{1:n})] - \log_{1/q} n|\right),
\nonumber
\end{IEEEeqnarray}
where expectations are with respect to $p^{\otimes n}.$
Since $p$ is Bernoulli($q$), then,
from the definition of $R(\cdot),$
$\{\mathbb{E}[R(X_{1:n})]\}_{n\in\mathbb{N}}$ and $\{\mathrm{Var}(R(X_{1:n}))\}_{n\in\mathbb{N}}$ are sequences of polynomials in $q$ with computable sequences of coefficients --- indexed by $n$ and exponent.
Since $q$ is computable, then, $\{\mathbb{E}[R(X_{1:n})]\}_{n\in\mathbb{N}}$ and $\{\mathrm{Var}(R(X_{1:n}))\}_{n\in\mathbb{N}}$ are computable
sequences 
(of real numbers).
Thus, from
part (ii) of
Lemma~\ref{lemma:semicomputability_preserved_by_certain_operations},
$\{a_n\}_{n\in\mathbb{N}}$
is a
computable
sequence (of reals)
and $\delta_{\textrm{run}}$ is computable.
Then,
$
\mathbb{E}_{p^{\otimes n}}
[2^{\delta_{\textrm{run}}(X_{1:n})}]
$
rewrites as
\begin{IEEEeqnarray}{rCl}
\IEEEeqnarraymulticol{3}{l}{
2^{-a_n} \left(
1
+ \mathbb{E}\big[|R(X_{1:n}) - \log_{1/q} n|\big]
- \big|\mathbb{E}[R(X_{1:n})] - \log_{1/q} n\big|
+ \big|\mathbb{E}[R(X_{1:n})] - \log_{1/q} n\big|
\right)
}
\nonumber\\
&\le&
2^{-a_n}
\left(
1
+ \sqrt{
\mathbb{E}\big[|R(X_{1:n}) - \log_{1/q} n|\big]^2
- \big|\mathbb{E}[R(X_{1:n})] - \log_{1/q} n\big|^2
}
+ \big|\mathbb{E}[R(X_{1:n})] - \log_{1/q} n\big|
\right)
\nonumber\\
& \le &
2^{-a_n}
\left(
1
+ \sqrt{
\mathbb{E}\big[\big(R(X_{1:n}) - \log_{1/q} n\big)^2\big]
- \big|\mathbb{E}[R(X_{1:n})] - \log_{1/q} n\big|^2
}
+ \big|\mathbb{E}[R(X_{1:n})] - \log_{1/q} n\big|
\right)
=
1\nonumber
\end{IEEEeqnarray}
where
the first bound follows from
$0 \; {\leq} \; \beta \; {\leq} \; \alpha
{\implies}
\alpha \; {-} \; \beta \leq \sqrt{\alpha^2 \; {-} \;\beta^2},$
and
Jensen's inequality for $u \mapsto |u|,$
and
the second
from
Jensen's inequality for
$u \mapsto u^2.$
It only remains to show that $\{a_n\}_{n\in\mathbb{N}}$ is bounded.
This is a straightforward consequence of the 
asymptotic characterization of $\mathbb{E}[R(X_{1:n})]$ and $\mathrm{Var}(R(X_{1:n}))$ due to Gordon, Shilling
and Waterman~\cite[Theorem~2]{GordonRunsPTRF} --- see also the introduction of \cite{GordonRunsPTRF} for a high-level discussion---, building on earlier results of Guibas and Odlyzko~\cite{GuibasLongPatternsZWVG}
and Boyd~\cite{BoydLosingRunsUnpublished}.


\section{On the total variation distance
}\label{app:TV_small_batch}

\subsection{Some lemmas
}

The following lemmas correspond to Lemmas V.1, V.2, and Equation (29) in \cite{2013PaulCuffDistributedChannelSynthesis}.

\begin{lemma}\label{lemma:TV_joint_to_TV_marginal}
Consider two
finite sets $\mathcal{W}$ and $\mathcal{L}.$ 
Let $\Pi$ and $\Gamma$ be two distributions on
$\mathcal{W} \times \mathcal{L}.$ Then,
$
\| \Pi_W - \Gamma_W \|_{TV} \leq \| \Pi_{W,L} - \Gamma_{W,L} \|_{TV}.
$
\end{lemma}
\begin{lemma}\label{lemma:TV_same_channel}
Consider two
finite sets $\mathcal{W},\mathcal{L}$
and
two distributions
$\Pi
,
\Gamma$
on
$\mathcal{W} \times \mathcal{L}.$ Then when using the same conditional probability kernel $\Pi_{L|W},$ we have
$
\| \Pi_W \Pi_{L|W} - \Gamma_W \Pi_{L|W} \|_{TV} = \| \Pi_W - \Gamma_W \|_{TV}.
$
\end{lemma}
\begin{lemma}\label{lemma:continuity_TV}
Consider a
finite set $\mathcal{W}.$
Let $\Pi$ and $\Gamma$ be two distributions on
$\mathcal{W},$ and $f:\mathcal{W} \to \mathbb{R}$ be a bounded function. Then,
$
| \ \mathbb{E}_{\Pi}[f] - \mathbb{E}_{\Gamma}[f] \ |
\leq
2\max(|f|) \cdot \|\Pi-\Gamma\|_{TV}.
$
\end{lemma}

\subsection{Proof of Claim \ref{claim:TV_on_small_batch_one_shot}}

Fix two distributions $P
,
Q$
on
$\mathcal{X}$
and some
$B
\in\mathbb{N}
.$
Then, we have, with the convention $\Pi \otimes \Gamma^{\otimes 0} \equiv \Pi,$
\begin{IEEEeqnarray}{rCl}
\|P^B-Q^B\|_{TV} &=& \Bigg\|\sum_{k{=}1}^B (P^{\otimes (B-k+1)}\otimes Q^{\otimes(k-1)}-P^{\otimes(B-k)} \otimes Q^{\otimes k})\Bigg\|_{TV}
\nonumber\\
&\leq& \sum_{k{=}1}^B \|P^{\otimes (B-k+1)}\otimes Q^{\otimes(k-1)}-P^{\otimes(B-k)} \otimes Q^{\otimes k}\|_{TV}
\label{eq:in_proof_telescoping_using_triangle_ineq}\\
&\leq& \sum_{k{=}1}^B \|
P^{\otimes (B-k)}
\otimes P
\otimes Q^{\otimes(k-1)} -
P^{\otimes(B-k)}
\otimes Q
\otimes
Q^{\otimes (k-1)}
\|_{TV}
\nonumber\\
&\leq& \sum_{k{=}1}^B \|P-Q\|_{TV} = B \|P-Q\|_{TV},\label{eq:in_proof_telescoping_using_TV_same_channel}
\end{IEEEeqnarray}
where \eqref{eq:in_proof_telescoping_using_triangle_ineq} follows from the triangle inequality for the
TVD
; and \eqref{eq:in_proof_telescoping_using_TV_same_channel} follows form Lemma \ref{lemma:TV_same_channel} with $W=X_{B-k+1},$ $\Pi_W \equiv P$ and $\Gamma_W \equiv Q$ --- which applies since $\mathcal{X}$ is finite.


\section{Existence of a universal $p^{\otimes *}$-critic}\label{app:proof_of_existence_universal_critic}

We provide a proof of Theorem \ref{thm:universal_semi-measure_and_critic}.
From \cite[Theorem~4.3.1]{BookKolmogorovComplexity}, there exists a sequence $\{q_n\}_{n \geq 1}$ containing all lower semi-computable semi-measures on $\{0,1\}^*,$ and a sequence $\{\pi_n\}_{n \geq 1}$ of (strictly) positive reals, such that the mixture defined by
\begin{equation}\label{eq:mixture}
\mathbf{m} := \sum_{n \geq 1} \pi_n q_n
\end{equation}
is a lower semi-computable semi-measure on $\{0,1\}^*.$
For every $n\in\mathbb{N},$ let $\mathbf{m}(\mathcal{X}^n)
:=
\sum_{x_{1:n}\in\mathcal{X}^n}\mathbf{m}(x_{1:n}).$
From \eqref{eq:mixture}, we have $\forall x \in \{0,1\}^*, \mathbf{m}(x)>0.$ Moreover, $\forall x_0 \in \mathcal{X}, \ p(x_0)>0,$ thus $\forall x \in \cup_{n \in \mathbb{N}} \mathcal{X}^n,$ $p^{\otimes *}(x)>0.$ Define the function $\delta_0,$ by
\begin{IEEEeqnarray}{c}
\forall n \in\mathbb{N}, \forall x_{1:n}\in\mathcal{X}^n, \
\delta_0(x_{1:n}):= \log\Big(\dfrac{\mathbf{m}(x_{1:n})}{\mathbf{m}(\mathcal{X}^{n})p^{\otimes n}(x_{1:n})}\Big).
\label{eq:def_delta_0}
\end{IEEEeqnarray}
Then, for any $n\in\mathbb{N},$ we have
\begin{IEEEeqnarray}{c}
\sum_{x\in
\mathcal{X}
^n
}
p^{\otimes n}
(x
)2^{\delta_0(x)}
=
\sum_{x\in
\mathcal{X}
^n
}
\dfrac{\mathbf{m}(x)}{\mathbf{m}(\mathcal{X}^{n})}
\leq 1.
\label{eq:delta_0_is_a_critic_according_to_our_def}
\end{IEEEeqnarray}
Hence, $\delta_0$ is a $p^{\otimes *}$-critic.
Fix a lower semi-computable $p^{\otimes *}$-critic $\delta.$ Define the map $q_\delta : \{0,1\}^* \to \mathbb{R}$ by
\begin{IEEEeqnarray}{c}
\forall x \in \cup_{n \in \mathbb{N}} \mathcal{X}^n, \quad q_\delta(x) := \mathbf{m}(\mathcal{X}^{l(x)})2^{\delta(x)}p^{\otimes *}(x),
\label{eq:in_proof_existence_of_universal_critic_def_q_delta}
\end{IEEEeqnarray}
and $x\mapsto 0$ elsewhere.
From Lemma \ref{lemma:semicomputability_preserved_by_certain_operations} (iii),
the function which is null outside of $\cup_{n \in \mathbb{N}}\mathcal{X}^n,$ and defined by $x \mapsto \mathbf{m}(\mathcal{X}^{l(x)})$ on $\cup_{n \in \mathbb{N}}\mathcal{X}^n,$ is lower semi-computable.
Moreover, $x \mapsto 2^{\delta(x)}$ and $x \mapsto p^{\otimes *}(x)$ are lower semi-computable by Lemma \ref{lemma:semicomputability_preserved_by_certain_operations} (i) and (iii) respectively.
Thus, $q_\delta$ is the product three non-negative lower semi-computable functions.
Hence, $q_\delta$ is lower semi-computable by Lemma \ref{lemma:semicomputability_preserved_by_certain_operations} (i).
Moreover,
\begin{IEEEeqnarray}{rCl}
\sum_{x \in \{0,1\}^*} q_\delta(x)
=
\sum_{n \in \mathbb{N}} \mathbf{m}(\mathcal{X}^n) \sum_{x \in \mathcal{X}^n} 2^{\delta(x)}p^{\otimes n}(x)
\leq
\sum_{n \in \mathbb{N}} \mathbf{m}(\mathcal{X}^n)
& \leq & 1,
\nonumber
\end{IEEEeqnarray}
where
the first bound
follows from the definition of a $p^{\otimes *}$-critic; and
the second
follows
from the fact that $\mathbf{m}$ is a semi-measure.
Therefore, $q_\delta$ is a lower semi-computable semi-measure.
Thus, from \eqref{eq:mixture}, we have $\mathbf{m} \geq \pi_{q_\delta} q_\delta,$ for some positive real $\pi_{q_\delta}.$
In order to derive \eqref{eq:property_defining_universal_critics}, fix $x \in \cup_{n \in \mathbb{N}}\mathcal{X}^n.$
From \eqref{eq:mixture}, we have $\mathbf{m}(x)>0.$ Therefore, since $\forall x_0 \in \mathcal{X}, p(x_0)>0,$ we have $q_\delta(x)>0.$
Thus, from \eqref{eq:def_delta_0}, we have
\begin{IEEEeqnarray}{rCl}
\delta_0(x)
=
\log\Big(\dfrac{\mathbf{m}(x_{1:n})}{\mathbf{m}(\mathcal{X}^{n})p^{\otimes n}(x_{1:n})}\Big)
\geq
\log\Big(\dfrac{\pi_{q_\delta}q_{\delta}(x_{1:n})}{\mathbf{m}(\mathcal{X}^{n})p^{\otimes n}(x_{1:n})}\Big)
=
\log(\pi_{q_\delta}) + \delta(x).
\nonumber
\end{IEEEeqnarray}
This is true for any lower semi-computable $p^{\otimes *}$-critic $\delta,$ and any $x \in \cup_{n \in \mathbb{N}}\mathcal{X}^n.$ Since $\log(\pi_{q_\delta})$ does not depend on $x,$ then property \eqref{eq:property_defining_universal_critics} holds. This concludes the proof.
\begin{remark}\label{rem:comparing_universal_critics_from_different_definitions}
As detailed in Remark \ref{rem:our_def_of_critic_differs_from_classical_definitions}, our definition of a $p^{\otimes *}$-critic is an adaptation of classical definitions \cite[Definitions~2.4.1~\&~4.3.8]{BookKolmogorovComplexity}.
In \cite[Definition~4.3.8]{BookKolmogorovComplexity}, a computable distribution $\{\pi(n)\}_{n\in\mathbb{N}}$ with $\pi(n)>0$ for all $n$ is assumed to be given.
The universal critic
under that definition is given by
\begin{IEEEeqnarray}{c}
\forall n \in\mathbb{N}, \forall x_{1:n}\in\mathcal{X}^n, \
\delta_1(x_{1:n}):= \log\Big(\dfrac{\mathbf{m}(x_{1:n})}{\Tilde{p}(x_{1:n})}\Big),
\nonumber
\end{IEEEeqnarray}
\cite[Theorem~4.3.5]{BookKolmogorovComplexity},
where $\Tilde{p}(x_{1:n}) = \pi(n)p^{\otimes n}(x_{1:n}).$ In other words, for any $\delta$ that is a $p^{\otimes *}$-critic according to \cite[Definition~4.3.8]{BookKolmogorovComplexity}, there exists a constant $c_\delta$ such that
\begin{IEEEeqnarray}{c}
\forall x \in \bigcup_{n \in \mathbb{N}} \mathcal{X}^n, \quad \delta_1(x) \geq \delta(x) - c_\delta.
\label{eq:def_universality_according_to_sum_test}
\end{IEEEeqnarray}
One can verify that 
this domination also holds for
all functions $\delta$ which are $p^{\otimes *}$-critics in the sense of Definition \ref{def:sum_critic_at_beginning_of_paper}, irrespective of the weights $\{\pi(n)\}_n.$ However, $\delta_1$ is not itself a critic according to Definition \ref{def:sum_critic_at_beginning_of_paper}
for all choices of $\{\pi(n)\}_n$
(Eq. \ref{eq:delta_0_is_a_critic_according_to_our_def} may fail to hold).
\end{remark}
\begin{remark}\label{remark:careful_not_to_forget_normalization}
The presence of weights $\{\mathbf{m}(\mathcal{X}^n)\}_n$ in the definition of $\delta_0$ (Eq. \ref{eq:def_delta_0}), or any other weights, as considered in Remark \ref{rem:comparing_universal_critics_from_different_definitions} and references therein, is natural: this ensures that the denominator is a semi-measure over $\cup_{n \in \mathbb{N}} \mathcal{X}^n,$ similarly to the numerator.
Moreover, such scaling is crucial for the critic to be universal.
Indeed,
consider the function $\delta_2$ that is a variant of $\delta_0$ with no scaling, and is defined by
\begin{IEEEeqnarray}{c}
\forall n \in\mathbb{N}, \forall x_{1:n}\in\mathcal{X}^n, \
\delta_2(x_{1:n}):= \log\Big(\dfrac{\mathbf{m}(x_{1:n})}{p^{\otimes n}(x_{1:n})}\Big).
\nonumber
\end{IEEEeqnarray}
One can readily prove that
$\delta_2$ is lower semi-computable
by using similar computability arguments as for $q_\delta$ (Eq. \ref{eq:in_proof_existence_of_universal_critic_def_q_delta}).
Moreover, since $\mathbf{m}$ is a semi-measure, we have $\delta_2\leq\delta_0,$ thus, from \eqref{eq:def_delta_0}, $\delta_2$ is also a $p^{\otimes *}$-critic.
One can readily show that
if
$\delta_2$
were universal, then it would
also 
be
a
universal
$p^{\otimes *}$-critic according to
\eqref{eq:def_universality_according_to_sum_test}
and
\cite[Definition~4.3.8]{BookKolmogorovComplexity}
---
for any given computable distribution $\{\pi(n)\}_{n\in\mathbb{N}}$ ---,
under which $\delta_1$ is universal (Remark \ref{rem:comparing_universal_critics_from_different_definitions}).
This would imply that
$|\delta_1 - \delta_2|$ is
bounded,
which is false by construction.
Hence,
$\delta_2$ is not a universal $p^{\otimes *}$-critic according to Definition \ref{def:sum_critic_at_beginning_of_paper}.
To wit, the theoretical concept of universality is only relevant when considering critics defined on a countably infinite set --- such as $\{0,1\}^*$ or $\cup_{n \in \mathbb{N}} \mathcal{X}^n.$
Hence, scaling such as considered here may be irrelevant in scenarios where critics are only defined on a finite set.
\end{remark}


\section{Additional semi-computability arguments}\label{app:proof_practical_semicomputability_lemma}

We provide a proof of the fact that $f \mapsto 2^f/(f+3)^2$
preserves lower semi-computability. The remainder of
Lemma \ref{lemma:semicomputability_preserved_by_certain_operations}
follows from similar arguments.
Let $f: \mathcal{E} \to \mathbb{R}$ be lower semi-computable
and
$(x,k) \mapsto \varphi_{f}(x,k)$
be
a computable function from $\mathcal{E}$ to $\mathbb{Q},$
monotonically
approaching $f$ from below
(Def.~\ref{def:semi_computability}).
There exists a real $\varepsilon \in (0,1)$ such that $u \mapsto 2^u/(3+u)^2$ is non-decreasing on $(-\varepsilon, \infty).$
Fix $x\in\mathcal{E}.$ Compute $k_1(x),$ the smallest
$k
\in\mathbb{N}
$ such that $\varphi_{f}(x,k) >-\varepsilon.$ For all
$k \in [1,k_1(x))\cap\mathbb{N},$ define $\varphi_{2^f/(3+f)^2}(x,k) :=0.$ Fix
$k\geq k_1(x).$ Let $a \in \mathbb{Z}$ and $b \in \mathbb{N}$ such that $\varphi_{f}(x,k)=a/b.$
Compute the greatest
$m
\in\mathbb{N}
$ such that $(m/2^k)^{b} \leq 2^a/(3+a/b)^{2b}.$
Then, set $\varphi_{2^f/(3+f)^2}(x,k) := m/2^k.$
Therefore,
\begin{equation}\label{eq:tightening_gap_in_proof_semicomputability_exponential_times_inverse}
\forall k \geq k_1(x), \ \ 0 \leq
\dfrac{2^{\varphi_{f}(x,k)}}{(3+\varphi_{f}(x,k))^2} - \varphi_{2^f/(3+f)^2}(x,k)
< \scalebox{0.9}{$\dfrac{1}{2^k
}$}.
\end{equation}
Note that $\varphi_{2^f/(3+f)^2}$ is defined via an algorithm that halts for every input $(x,k)$ and outputs a rational number. Thus,
it is a computable rational-valued function.
From \eqref{eq:tightening_gap_in_proof_semicomputability_exponential_times_inverse}, and since $k \mapsto \varphi_{f}(x,k)$ is non-decreasing, and $u \mapsto 2^u/(3+u)^2$ is non-decreasing on $(-\varepsilon, \infty),$ we have
\begin{equation*}
\forall k \geq k_1(x)+1, \ \ \varphi_{2^f/(3+f)^2}(x,k-1) \leq \dfrac{2^{\varphi_{f}(x,k)}}{(3+\varphi_{f}(x,k))^2}.
\end{equation*}
Since $\varphi_{2^f/(3+f)^2}(x,k-1)$ can also be written in the form $m'/2^k,$ then, from the maximality of the integer $m$ appearing in the construction of $\varphi_{2^f/(3+f)^2}(x,k),$ we have 
\begin{equation}\label{eq:monotonicity_in_proof_semicomputability_exponential_times_inverse}
\forall k \geq k_1(x)+1, \ \ \varphi_{2^f/(3+f)^2}(x,k-1) \leq \varphi_{2^f/(3+f)^2}(x,k).
\end{equation}
This is also true for all integers $k \in [2,k_1(x)+1).$
Properties \eqref{eq:tightening_gap_in_proof_semicomputability_exponential_times_inverse} and \eqref{eq:monotonicity_in_proof_semicomputability_exponential_times_inverse} imply
that $2^f/(3+f)^2$ is lower semi-computable.




\ifCLASSOPTIONcaptionsoff
  \newpage
\fi



\bibliographystyle{IEEEtran}
\bibliography{biblio}
\end{document}